\documentclass[preprint2]{proto}
\usepackage{times}
\usepackage{color}

\newcommand{\msun}{{\,\rm M}_{\odot}}

\newcommand{\kms}{\mathrm{km.s}^{-1}}
\newcommand{\tco}{\ifmmode {^{13}{\rm CO}} \else {$^{13}{\rm CO}$}\fi}
\newcommand{\dco}{\ifmmode {^{12}{\rm CO}} \else {$^{12}{\rm CO}$}\fi}
\newcommand{\cdo}{\ifmmode {{\rm C}^{18}{\rm O}} \else {${\rm C}^{18}{\rm O}$}\fi}
\newcommand{\juz}{\ifmmode {{\rm J}=1\rightarrow\0} \else
{J=1$\rightarrow$0}\fi}
\newcommand{\jdu}{\ifmmode {{\rm J}=2\rightarrow\1} \else
{J=2$\rightarrow$1}\fi}
\newcommand{\jtd}{\ifmmode {{\rm J}=3\!\rightarrow\!2} \else
{${\rm J}=3\!\rightarrow\!2$} \fi}
\newcommand{\jcq}{\ifmmode {{\rm J}=5\!\rightarrow\!4} \else
{${\rm J}=5\!\rightarrow\!4$} \fi}

\newcommand{\dbf}{}
\newcommand{\bbf}{}

\begin{document}

\title{\textbf{\LARGE Physical and chemical structure of planet-forming disks probed by millimeter observations and
modeling}}

\author {\textbf{\large Anne Dutrey$^1$, {\dbf Dmitry} Semenov$^2$, Edwige Chapillon$^3$, Uma Gorti$^4$, St\'ephane
Guilloteau$^1$, Franck Hersant$^1$, Michiel Hogerheijde$^5$, Meredith Hughes$^6$,  Gwendolyn Meeus$^7$, Hideko Nomura$^8$, Vincent
Pi\'etu$^9$, Chunhua Qi$^{10}$, Valentine Wakelam$^1$}}
\affil{\small\em $^1$ Laboratoire d'Astrophysique de Bordeaux, CNRS, University of Bordeaux, France; $^2$Max-Planck-Institut f\"ur Astronomie, K\"onigstuhl 17, 69117 Heidelberg, Germany;
$^3$Institute of Astronomy and Astrophysics, Academia Sinica, P.O. Box 23-141,
Taipei 106, Taiwan, Republic of China;$^4$SETI Institute, 189 Bernardo Ave, Mountain View, CA 94043 USA -
NASA Ames Research Center, MS 245-3, Moffett Field, CA 94035 USA;$^5$Leiden Observatory, Leiden University, P.O. Box 9513, 2300 RA, Leiden, The Netherlands;$^6$Wesleyan University Department of Astronomy, Van Vleck Observatory, 96 Foss Hill Drive, Middletown, CT 06459, USA;$^7$UAM, Dpto. Física Teórica, Mód.15, Fac. de Ciencias, 28049, Madrid, Spain;$^8$Department of Astronomy, Graduate School of Science, Kyoto University, Kyoto 606-8502, Japan ;$^9$IRAM 300 rue de la Piscine, Domaine Universitaire 38406 Saint Martin d'Hères, France;$^{10}$Harvard-Smithsonian Center for Astrophysics
Cambridge, MA 02138, USA.}

\begin{abstract}
\baselineskip = 11pt
\leftskip = 0.65in
\rightskip = 0.65in
\parindent=1pc
{\small
Protoplanetary disks composed of dust and gas are ubiquitous around young stars and are commonly
recognized as nurseries of planetary systems. Their lifetime, appearance, and structure are determined by an interplay between stellar
radiation, gravity, thermal pressure, magnetic field, gas viscosity, turbulence, and rotation. Molecules and
dust serve as major heating and cooling agents in disks. Dust grains dominate the disk opacities, reprocess most of the
stellar radiation, and shield molecules from ionizing UV/X-ray photons.
Disks also dynamically evolve by building up planetary systems which drastically change
their gas and dust density structures.
Over the past decade significant progress has been achieved in our understanding of disk chemical
composition thanks to the upgrade or advent of new millimeter/Infrared facilities (SMA, PdBI, CARMA, Herschel, {\dbf e-VLA,
ALMA).}
Some major breakthroughs in our comprehension of
the disk physics and chemistry have been done since PPV.
This review will present and discuss the impact of such improvements on our understanding of
the disk physical structure and chemical composition.
\\~\\~\\~}
\end{abstract}

\section{\textbf{INTRODUCTION}}
\label{sec:intro}
The evolution of the gas and dust in protoplanetary disks is a key element that regulates the efficiency, diversity
and timescale of planet formation. The situation is complicated by the fact that the
dust and gas physically and chemically interact atop of the disk structure
that evolves with time. Initially, small dust grains are dynamically well coupled
to the gas and are later assembled by grain growth in bigger cm-sized particles which settle towards the disk midplane (see
Chapter by Testi et al.).
After large grains become dynamically decoupled from the gas, they {\dbf become} subject to head wind from the gas orbiting at
slightly lower
velocities and spiral rapidly inwards or experience mutual {\dbf destructive collisions}.
Collisionally-generated
small grains are either swept out by larger grains or stirred by turbulence into the disk atmosphere.
As a result, the dust-to-gas {\dbf ratio} and average dust {\dbf sizes} vary through the disk.
All these processes affect the disk thermal and density structures and thus {\dbf disk} chemical composition.

In the dense disk midplane, thermal equilibrium between gas and dust is achieved, with dust
transferring heat to gas by rapid gas-grain collisions. The disk midplane is well shielded from high-energy
stellar
radiation and thus is "dark", being heated indirectly via infrared emission from the upper layers, and have low
ionization degree and low turbulence velocities. {\dbf The ratio of
ions to neutral molecules determines the level of turbulence and regulates the redistribution of the angular momentum}.  The
outer disk
midplane is so cold that many gaseous molecules {\dbf are frozen} out onto dust grains, leading to the
formation of icy mantles that are steadily processed by cosmic rays.

Above the midplane a warmer, less dense region is located, where gas-grain collisional
coupling can no longer be efficient, and dust and gas temperatures  start to depart from each other, with
the gas temperature being usually higher. The intermediate disk layer is only partly shielded from the
ionizing radiation by the dust and thus is more ionized and dynamically active than the midplane.
{\dbf Rich gas-phase and gas-grain chemistries} enable synthesis of {\dbf many}  molecules in this so-called molecular
layer.

Finally, in the heavily irradiated, hot and dilute atmosphere
only simple atoms, ions, photostable radicals and PAHs are able to survive. {\dbf The global chemical evolution is dominated
by a limited set of gas-phase reactions in this disk region}.

{\bbf
Observations of protoplanetary disks of various ages and sizes surrounding Sun-like and intermediate-mass
stars help to resolve some of the related ambiguities. Detailed studies
of protoplanetary disks remain an observationally challenging task though, as disks are compact objects
and have relatively low masses.  At visual and infrared wavelengths disks are typically opaque, and one
uses (sub-)millimeter imaging with single-dish telescopes and antenna arrays to peer
through their structure. Since observations of the most dominant species in disks, H$_2$, are
impossible (except of the hot upper layers via its weak quadrupole IR transitions or of the inner disk via the fluorescent $FUV$ lines),
other molecules are employed to trace disk kinematics, temperature, density, and chemical structure.
The poorly known properties such as (sub-)millimeter dust emissivities and
dust-to-gas ratio make it hard to derive a total disk mass from dust emission alone.
Apart from a handful of simple molecules, like CO, HCO$^+$, H$_2$CO,
CS, CN, HCN and HNC, the molecular content of protoplanetary disks remains largely unknown.
}

{\bbf In the last ten years, upgraded and new millimeter (mm) or submillimeter (submm) facilities (IRAM, CARMA, SMA, Herschel and
ALMA)
have permitted the detection of a few other molecular species (DCN, N$_2$H$^+$, H$_2$O, HC$_3$N, C$_3$H$_2$, HD)
at better spatial and frequency resolution.
In the meantime, disk models have benefited from improvements in astrochemistry databases (like KIDA and UDFA'12), development of
coupled
thermo-chemical disk physical models, line radiative transfer codes, and better analysis tools.
The spatial distribution of molecular abundances in disks is still poorly determined, hampering a
detailed comparison with existing chemical models. Due to the complexity of the molecular line excitation,
unambiguous interpretation of the observational results necessitates advanced modeling of the disk
physical structure and evolution, chemical history, and radiative transfer.}

{\bbf
\begin{figure*}
 \epsscale{1.5}
\includegraphics[angle=270.0,width=16.0cm]{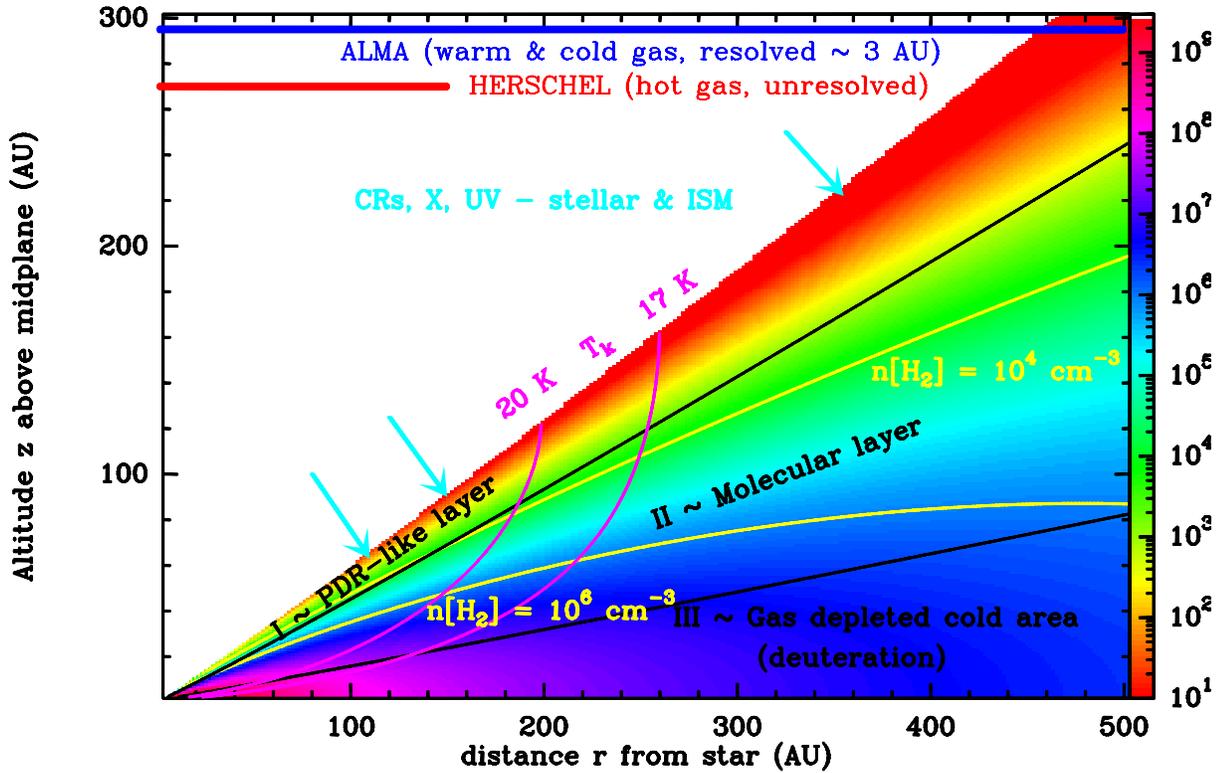}
 \caption{\small  Molecular disk: state of art after PPV. The model is
 done for a disk orbiting a young star of one solar mass.
 The volume density as a function of radius r and altitude z above the midplane is
 given in color. The location of the isotherms at 17 and 20 K are also shown.
 I,II and III correspond to the location of the three important layers of gas:
 the {\dbf PDR-like atmosphere irradiated by the star, the UV interstellar field and
 the cosmic rays (I), the molecular layer typically located between one and three scale heights (II),
 and the midplane where only molecular hydrogen and deuterated isotopologues of simple
 molecules are abundant in the gas phase (III). Top lines: the red line gives the area sampled by Herschel
 observations (unresolved). The blue line corresponds to the area sampled by ALMA (with the best resolution
 for a source at 150~pc).}}
 \label{state1}
 \end{figure*}
Figure \ref{state1} illustrates the state of art prior to the Protostars and Planets~VI
conference hold in Heidelberg in June 2013.
The disk structure (density and temperature) has been calculated for a {\dbf T~Tauri
star with the mass of $1 \msun$}. With the exception of the very inner disk ($R \leq 10$~AU),
where {\dbf dissipation of accretion energy} can be a source of {\dbf mechanical heat, disks} are mostly heated by the
stellar {\dbf radiation, including UV and X rays}, the interstellar UV field and the cosmic rays. Both the gas and
dust have vertical and radial temperature {\dbf gradients}, the upper {\dbf disk} layers being super-heated by
the {\dbf central star(s)}.
As grains grow and rapidly settle towards the
disk midplane (see Chapter by Testi et al.), the vertical temperature gradient {\dbf changes} and
the midplane further cools off.
This may have impact on the location {\dbf and size of the freeze-out zones of various molecules, where they mainly remain
bound to the dust grains}.

In summary, there are three different chemical zones {\dbf in disks: (1) the disk upper layer that
is similar to a dense PDR, which consist of simple ions, neutral species, and small dust grains, (2)
the warm intermediate molecular layer with many  molecules (including some complex ones) in the gas phase and rapid gas-dust
interactions, and (3) the cold midplane devoid of many gaseous molecules but icy-rich at radii $\ga 20-200$~AU,
similar to dense prestellar cores.}

{\dbf This review will present the recent improvements on our understanding of disk physical structure and
chemical composition, and their evolution with time, both from the observational but also theoretical perspectives.
Other recent comprehensive reviews on the subject of disk physics and chemistry are those by \citet{Bergin_09},
\citet{2010arXiv1011.4770S} and
\citet{2013arXiv1310.3151H}.}

\bigskip
\centerline{\textbf{2. DATA ANALYSIS AND BIAS}}
\bigskip

The physical structure of protoplanetary disks is {\dbf so unique because it has both the} strong density
and temperature gradients on relatively short spatial scales. For example, at a radius of 100 AU, the
midplane temperature is expected to be around {\dbf 7-12 K, while at the 1-3 scale heights (H) above (at about 15-50 AU),
the temperature is warm enough to facilitate} the presence of a molecular layer. In the
meantime, {\dbf the gas density drops by about one to four orders of magnitude.
As a consequence, the chemical conditions change rapidly} along the
line of sight, as {\dbf well as} the excitation conditions, leading in some cases to non-LTE effects.
Last but not least, a proper analysis of any line detection in disks requires an adequate handling of the line
formation process, {\dbf including possible non-LTE excitation, but also gas kinematics, } in particular, the disk Keplerian
shear \citep{2006A&A...448L...5G,Dutrey_ea07}.

Line analysis falls in two main categories: inversion methods, which attempt
to retrieve the physical parameters (e.g. the excitation temperature, the molecular
surface density distributions, etc..) and their confidence intervals,
and forward modeling, which evaluate whether a given model can represent the observations.
In the radiative transfer codes used to analyze observations or predict mm/submm line emissions,
the physical parameters (density, gas temperature, scale height or turbulence) are usually
specified as explained below:

\paragraph{Temperature:} {\dbf Disk temperature} is treated in three main ways:
1) the gas temperature is assumed to be {\dbf isothermal in the vertical direction and to follow
a power law in the radial direction, $T_k(r) \propto r^{-q}$ \citep{DGS94}},
2) a vertical gradient is also included \citep{DDG03,2011ApJ...740...84Q,2012ApJ...757..129R},
{\dbf using simple parametrization of} dust disk models possibly with the use of 
a simplified approach to the radiative transfer problem \citep{CG97,DAea98},
or 3) the temperature is directly {\dbf calculated by the dust
radiative equilibrium modeling} \citep{Pinte_ea08,2009ApJ...701..260I}.
This last category of models has a number of hidden {\dbf and often not well constrained} parameters. The surface density
distribution must be specified, as well as
the dust properties (minimum and maximum radii, size distribution index, composition) to calculate
the dust absorption coefficients. These models can accommodate grain growth and even dust settling
{\dbf \citep[e.g.][]{2007prpl.conf..555D,Hasegawa+Pudritz_2010,2011ARA&A..49...67W}}.
The dust disk properties are usually adjusted to  fit the
the Spectral Energy Distribution (SED), sometimes simultaneously with
resolved dust maps in the mm/submm \citep{2009ApJ...701..260I,2013A&A...553A..69G}. 
A last refinement is to also evaluate the gas
{\dbf heating and cooling and to account for the differences between
the gas and dust temperature as a function of disk location
(this will be discussed in Section 4.1})}. 

\paragraph{Density:} Two types of surface density must be considered: the total (mass) surface density and
individual molecule distributions. Some parametric models allow to adjust the molecular distributions,
either as power law surface densities \citep{Pietu_ea07}, or more sophisticated variations, such
as piece-wise power law integrated abundance profiles \citep{Qi_ea08,Guilloteau_ea12a} atop of a
prescribed H$_2$ surface density. In forward modeling, models computing the thermal balance
(dust and/or gas) in general assume a power law (mass) surface density profile. Exponentially
tapered profiles have also been invoked to explain
the different radii {\dbf observed} in CO isotopologues \citep{Hughes_ea09}, but are not
widely used in forward modeling.

\paragraph{Scale height:}  The gas is usually assumed to be in hydrostatic equilibrium.
The scale height is prescribed as a power law or self-consistently calculated from the
temperature profile, but the {\dbf feedback}
of the temperature profile on the vertical density profile is neglected in most cases.
The scale height can be expressed as \citep[e.g.][]{DDG03,Dutrey_ea07b}:
\begin{equation}
H(r) = \sqrt{\frac{2kr^3T(r)}{GM_*m_0}} = \sqrt{\frac{2k}{m_0}}\frac{r}{V_{\rm K}(r)}\sqrt{T(r)} = \sqrt{2}c_{\rm s}/\Omega,
\label{eq:scale_height}
\end{equation}
where $T(r)$ is the kinetic temperature, $M_*$ is the stellar mass, $m_0$ is the mean molecular weight of the gas, $V_{\rm K}(r)$
is the Keplerian velocity, $c_{\rm s}$ is the sound speed, and $\Omega$ is the angular velocity.

Self-consistent calculation
of the hydrostatic equilibrium with vertical temperature gradients is however taken into
account in some models \citep[e.g.][]{DAea98, Gorti_Hollenbach04}.

\paragraph{Turbulence:} Turbulence is usually mimicked as by setting the local line
width $\Delta V = \sqrt{v_{th}^2+v_{turb}^2}$ where $v_{th}$ is the thermal broadening
and $v_{turb}$ the turbulent term \citep{DDG03,Pietu_ea07,Hughes_ea11a}.

\paragraph{Excitation conditions:}
The first rotational levels of most of the simple molecules observed so far with current
arrays such as CO, HCO$^+$, CN and HCN should be thermalized in the molecular
layer, given the expected H$_2$ densities. This strongly simplifies the analysis of the data.
This is {\dbf however} less true for higher levels (above J=3), as noted by \citet{Pavlyuchenkov_ea07a},
and a correct treatment of the non-LTE
excitation conditions is needed. These {\dbf non-LTE effects will become more common}
with submm ALMA (Atacama Large Millimetre Array) observations {\dbf sensitive to weaker lines and to
more warmer, inner disk regions with strong dust continuum background emission.}

\paragraph{Thermo-chemical models:} The most sophisticated approach is to
couple the calculation of the thermal and density structures to the derivation of
the chemistry to make molecular line predictions \citep{Woitke_ea09,2010A&A...510A..18K,ANDES}. However,
the computing requirements limit this kind of approach to forward modeling only. Despite
the interest of such models to understand the role of various processes, it is important
to remember that they also have their own limitations, such as the use of
equilibrium chemistry, the choice of dust properties, or an underlying surface
density profile. {\bbf We review these models in  Section 4.1.} 

\bigskip
\noindent
\textbf{2.1 Unresolved Data}
\bigskip

For most unresolved data, the SED at near-, mid-, and far-infrared
and (sub-)millimeter wavelengths is commonly used
to derive the disk structure from radii $<1$ AU all the way
to the outer radius {\dbf \citep{2007prpl.conf..555D}, as
the SED traces the distribution of the dust in the disk.
The dust is supported by the gas, so information on the gas radial
distribution and scale heights can also be derived. However, this assumes that the dust is
dynamically coupled to the gas, which is only true for small particles ($\la 100\mu$m).}
Complicating details are the presence of multiple
dust populations, vertical settling and radial drift of dust constituents,
generally poorly constrained dust opacities (see also Testi et al. chapter),
and the gas and dust thermal decoupling.

Moreover, the IR part of the SED is in general insensitive to
the total dust content, which is only constrained by the (sub-)mm
range. Neither of them is sensitive to the disk outer radius.
Spatially integrated, spectrally resolved line profiles {\dbf can  be
the tools} to sample disk properties, because the line formation
in a Keplerian disk links velocities to radial distances. This
can be used to obtain disk radii \citep{2013A&A...549A..92G} or evidence
for central holes \citep[][, see also Sec.5]{2008A&A...490L..15D}.

\bigskip
\noindent
\textbf{2.2 Resolved Interferometric Data}
\bigskip

Resolved interferometric molecular maps on nearby
protoplanetary disks are routinely obtained from most
mm/submm arrays such as the SMA (Submillimetre Array),
CARMA (Combined Array for Research in Millimeter-wave Astronomy)
and IRAM (Intitut de RadioAstronomie Millimetrique)
Plateau de Bure Interferometer (PdBI).

The UV coverage of these facilities is still a limiting factor
and data are generally compared to disk models by $\chi^2$
minimizations performed in the Fourier plane in order to avoid
non linear effects due to the deconvolution. This later step
should no longer be necessary for most ALMA configurations.

{\dbf Other important} limitations are due to the assumed density and temperature
laws. Current arrays do not have enough sensitivity and angular resolution
to allow a fine {\dbf determination} of the radial and vertical structure. Depending
on the way the temperature is calculated (from the dust) or determined (from
thermalized CO lines such the 1-0 or the 2-1 transitions), the biases are
different. In the first case, the gas temperature is directly dependent
on the poorly known dust properties, and there is no {\dbf direct}
gas temperature measurements. In the second case, an excitation
temperature is determined, but its interpretation as the kinetic
temperature depends on the robustness of the LTE hypothesis, and
the region sampled by the {\dbf measurement, as it is} depends on the chemical behavior
of the observed molecule.

\bigskip
\centerline{\textbf{3. MOLECULAR OBSERVATIONS}}
\bigskip

We discuss in this section how molecular observations obtained at mm/submm
wavelengths can constrain the disk structures. Some far-IR results obtained
mostly with Herschel are also discussed but the gas properties of the
warm surface and inner disk derived from IR observations (Spitzer, Herschel)
are discussed in the chapter by Pontoppidan et al.


\bigskip
\noindent
\textbf{3.1 Detected Species}
\bigskip

\begin{figure*}
 \epsscale{1.5}
\includegraphics[angle=0.0,width=16.0cm]{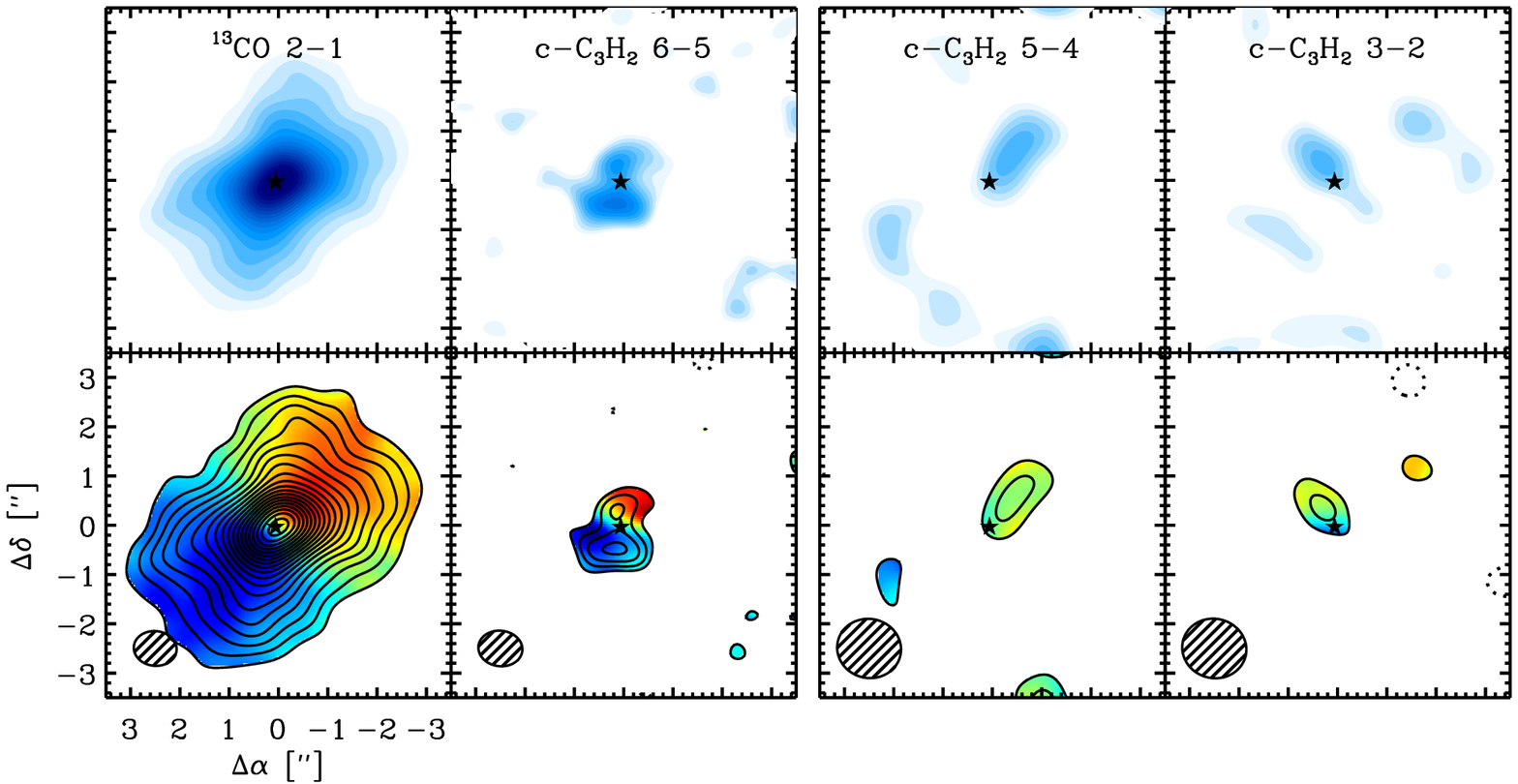}
 \caption{From \cite{2013ApJ...765L..14Q}. The integrated intensity maps and
and intensity-weighted mean velocity fields of $^{13}$CO 2-–1 and c$-$C$_3$H$_2$
6–-5 lines (left panels), as well as c-C$_3$H$_2$ 5-–4 and 3–-2 lines (right panels)
toward HD~163296. The resolved velocity field of the c$-$C$_3$H$_2$
6–-5 line agrees with the CO kinematics. In the c$-$C$_3$H$_2$
maps, the first contour marks 3$\sigma$
followed by the 1$\sigma$ contour increases.
The rms varies between 6 and 9 mJy.km/s per beam.
Synthesized beams are presented in the lower left corners.
The star symbol denotes the continuum (stellar) position.
The axes are offsets from the pointing center in arc-seconds.}
\label{hd163296}
\end{figure*}

Since H$_2$ cannot be used as a tracer of the bulk of gas mass, the study of the more abundant molecules
after H$_2$ is mandatory to improve our knowledge on the gas disk density and mass distribution.
After the detection of a few simple molecules in the T~Tauri disks surrounding
the 0.5 $\msun$ DM Tau and the binary system GG Tau (1.2 $\msun$) by \citet{DGG97}, 
millimeter/submmilliter facilities have observed several protoplanetary disks
around young stars with mass ranging between $\sim 0.3$ and $2.5 \msun$ \citep[e.g.][]{Kastner_ea97,Thi_ea04}.
The main result of mm/submm molecular studies is that detections are limited
to the most abundant molecules found in cold molecular clouds: {\dbf CO (with $^{13}$CO and C$^{18}$O), HCO$^+$ (with
H$^{13}$CO$^+$ and DCO$^+$), CS, HCN (with HNC and DCN),} CN, H$_2$CO, N$_2$H$^+$, C$_2$H and, very recently, HC$_3$N
\citep{Chapillon_ea12b}
 followed by
the detection {\dbf of cyclic C$_3$H$_2$ in the disk surrounding HD~163296 using ALMA
\citep[see Fig.\ref{hd163296};][]{2013ApJ...765L..14Q}}.
H$_2$O has also been detected by the  Herschel satellite in TW Hya and DM Tau (marginal detection)
\citep{Bergin_ea10a,2011Sci...334..338H}, 
while the main reservoir of {\bbf elemental deuterium}, HD, has been detected
in TW Hya  by \citet{2013Natur.493..644B}. 
The mm/submm molecular detections are summarized in Table \ref{tbl-1}.

\begin{deluxetable}{lccc}
\tabletypesize{\small}
\tablecaption{A table of detected species in
T~Tauri (cold) and Herbig Ae (warm) outer disks using mm/submm facilities
\label{tbl-1}}
\tablewidth{0pt}
\tablehead{Species &  T~Tauri cold disk & Herbig Ae warm disk & Telescope \\
& $T < 15$~K at $R>50$~AU & $T < 15$~K at $R>200$~AU \\}
\startdata
CO, $^{13}$CO, C$^{18}$O & Many& Many& Many   \\
CN & Many& Many & Many   \\
HCN, HNC & Several, DM Tau&  & IRAM, SMA \\
DCN & TW Hydra & - & SMA, ALMA  \\
CS & Several & Several & IRAM, SMA  \\
C$_2$H & Several & Several & IRAM, SMA  \\
H$_2$CO & Several & A few & IRAM, SMA  \\
HCO$^{+}$ & Several & Several & IRAM, SMA  \\
DCO$^{+}$& TW Hya, DM Tau & HD~163296 & IRAM, SMA \\
N$_2$H$^{+}$ & 3-4 &  HD~163296 & IRAM, SMA, ALMA \\
HC$_3$N & 3-4 & & IRAM \\
c-C$_3$H$_2$ & - & HD~163296 & ALMA \\
H$_2$O & TW Hya, DM Tau &  & Herschel  \\
HD & TW Hya&  & Herschel  \\
\enddata
\end{deluxetable}

\bigskip
\noindent
\textbf{3.2 Outer Disks Structure ($R > 30$ AU)}
\bigskip

\paragraph{CO as a tracer of the disk structure:}
Resolved spectro-imaging observations of CO isotopologues  have been
so far the most powerful method to constrain the geometry (outer radius, orientation and
inclination) and velocity pattern of protoplanetary disks \citep{Koerner_ea93,DGS94,GD98}.

As the first lowest rotational levels of CO are thermalized, the observations
of the J=1-0 and J=2-1 transitions permit a direct measurement of the gas
temperature and surface densities. Furthermore, because of their different opacities,
the  $^{12}$CO, $^{13}$CO J=1-0 and J=2-1
lines sample different disk layers, probing the temperature gradient as
a function of height \citep[see][for a detailed discussion]{Dutrey_ea07b}.
With the optically thicker lines (e.g. $^{12}$CO J=2-1), the temperature can be
measured in a significant disk fraction, while the estimate of the molecular
surface density is only possible in the outer optically thin region \citep{DDG03,Pietu_ea07}.

\paragraph{Temperature structure:}
Closer to the midplane, several studies on various molecular lines
such as CCH 1-0 and 2-1 \citep{Henning_ea10}, HCN 1-0 and CN 2-1 \citep{Chapillon_ea12a} and
CS 3-2 \citep{Guilloteau_ea12a} suggest that the gas is cold, with apparent
temperatures of the order
of $\sim 10-15$~K at 100 AU for T Tauri disks. As this is below the CO freeze-out
temperature (17 - 19 K), CO and most other molecules {\dbf should be severely
depleted from the gas phase (apart from H$_2$, H$_3^+$ and their deuterated isotopologues)}.
Disks around HAe stars appear warmer \citep{Pietu_ea07}.
To explain the low {\dbf molecular} temperatures observed in the T~Tauri disks, several
possibilities can be invoked. With the physical structure
predicted by standard disk models, the observed molecular transitions of CN, CCH, HCN and CS
should be thermalized in a large area of these disks.
Hence, a subthermal excitation for these is unlikely, especially as the derived low temperatures
are similar to those obtained from (thermalized) CO 1-0 and 2-1 transitions \citep{Pietu_ea07}.
\citet{Chapillon_ea12a} have also investigated the possibility
to have a lower gas-to-dust ratio (by about a factor $\sim 6$). In
that case subthermal excitation becomes possible but the predicted column densities
for HCN and CN are low compared to the observed ones.
Turbulence may play a role by transporting gaseous molecules into the cold midplane
regions on a shorther timescale
than the freeze-out timescale \citep{Semenov_ea06,Aikawa_07}, but may be insufficient
\citep{Hersant_ea09,Semenov_Wiebe11a}, see also
Sect.4.2.2.1.
Finally, another
possibility would be that ``cold'' chemistry for molecules such as CN or CCH, and
in particular photo-desorption rates, are poorly known
\citep{Oeberg_ea09a,Hersant_ea09,Walsh_ea12}.
Tracing the very cold midplane is challenging since most molecules
should be frozen onto grains. The best candidate to trace this zone are the H$_2$D$^+$ or D$_2$H$^+$ ions,
because of their smaller molecular masses, but they remain to be detected in protoplanetary disks (see Sect.~5.3
and Ceccarelli et al. chapter in this book).

Higher in the disk, higher transitions of CO permit to constrain the
vertical structure. Using CO 3-2 and 6-5 maps obtained with the SMA,
\citet{2004ApJ...616L..11Q,Qi_ea06} 
have investigated the temperature  and the
heating processes in the T~Tauri disk surrounding TW Hya. They found
{\dbf that} the intensity of CO 6-5 emission can only be explained in presence
of {\dbf additional heating by stellar} UV and X-rays {\dbf photons}.

While a significant fraction of T~Tauri disks exhibits low gas temperatures, several studies
of Herbig Ae disks found that these disks are warmer, as predicted by the thermo-chemical disk models.
\citet{Pietu_ea07} and
\citet{Chapillon_ea12a} {\dbf found from the} CO, HCN and CN studies with the PdBI that the
molecular disk surrounding MWC\,480 has a typical temperature of about 30 K at
radius of 100 AU around the midplane. In the case of HD\,163296, \citet{2011ApJ...740...84Q} 
used a multi-transition and
multi-isotopologues study of CO (from SMA and CARMA arrays) to determine the location CO snowline,
where the CO column density changes by a factor 100. They found a radius of $\sim$ 155~AU
of this snowline. This result, consistent with the SMA  observation of  H$_2$CO
\citep{2013ApJ...765...34Q}, is confirmed by the ALMA observation of DCO$^+$ \citep{2013arXiv1307.3420M}.
More recently, \citet{2013arXiv1307.7439Q} found that the CO snowline in TW Hya is likely
located at a radius $\sim$ 30 AU, where the inner edge of N$_2$H$^+$ ring is detected.
This N$_2$H$^+$ ring is the result of an increase in N$_2$H$^+$ column density in
the region where CO abundances are low, because CO efficiently destroys N$_2$H$^+$ by
the proton transfer reaction: N$_2$H$^+$ + CO $\rightarrow$ N$_2$ + HCO$^+$.

\paragraph{Molecular density structure:}

The chemical behavior and the associated abundance variations of the molecules used
to characterize the gas precludes an absolute determination of the gas mass.
Recently, the far-infrared fundamental rotational line of HD has been
detected in the TW~Hya disk with {\it Herschel} \citep{2013Natur.493..644B}. 
The abundance distribution of HD closely follows that of H$_2$.
Therefore, HD, which has a weak permanent dipole moment, can be used
as a direct probe of the disk gas mass. The inferred mass of the TW~Hya disk is $\ga 0.05M_\odot$,
which is surprising for a $3-10$ Myr old system. Other detections of HD using SOFIA
could be a very powerful way to directly constrain the gas mass, assuming the thermal
structure is reasonably known.

{\dbf Whereas} the absolute determination of the gas mass is not yet possible, the radial and
vertical distributions of the molecular gas are better known.
Several complementary studies of the gas and dust distributions assuming either power law
or exponential decay for the surface density distribution have been recently
performed \citep{2009ApJ...701..260I,Hughes_ea09,Hughes_ea11a,Guilloteau_ea11a}.
Most of these studies are based on the analysis of dust maps (see also chapter by
Testi et al.) {\dbf but these results are still limited in sensitivity and angular resolution.}

{\dbf The vertical location of
the molecular layer was investigated in several studies, but in an indirect} way. \cite{Pietu_ea07}
found that the apparent scale heights of CO are larger than
the expected hydrostatic value in DM\,Tau, LkCa\,15 and MWC\,480.
\citet{Guilloteau_ea12a} showed that the
CS 3-2 and 5-4 PdBI maps are best explained if the CS layer is located
about one scale height above the midplane, in agreement with model predictions.
However, the first {\em direct} measurement of the location of the molecular
layer has only been {\dbf recently} obtained with ALMA observations of HD\,163296, where the CO
emission clearly originates from {\dbf a layer at $\sim 15^\circ$ above the disk plane
\citep{Rosenfeld+etal_2013,2013A&A...557A.133D}, at a few hundred AU.}

\paragraph{Gas-grain coupling:}
{\dbf Several observational studies indicate} the importance of the
grain surface chemistry for disks.

Using the IRAM 30-m telescope, \citet{Dutrey_ea11a} failed to detect SO and H$_2$S in DM Tau,
GO Tau, LkCa15 and MWC480 {\dbf disks. They} compared the molecular column densities derived from the observations with chemical
predictions made
using Nautilus \citep{Hersant_ea09} and the density and temperature profiles taken from
previous analysis \citep[e.g.][]{Pietu_ea07}. They reproduced the SO upper limits and CS column densities
reasonably well, but failed to match the upper limits obtained on H$_2$S by at least one order of magnitude.
This suggests that at the high densities and low temperatures encountered around disk midplanes,
H$_2$S remains locked onto the grain surfaces {\dbf and where it also gets destroyed to form other species}. Indeed, some recent
experiments by \citet{Garozzo_ea10} have shown
that H$_2$S on grains is easily destroyed by cosmic rays and lead to the formation of
C$_2$S, SO$_2$ and OCS on grains. These studies also suggest that most of the sulphur may be in
the form of a sulphur rich residuum, which could be polymers of sulfur or amorphous aggregates
of sulfur \citep{Wakelam_ea04a}. The associated grain surface reactions are
not yet incorporated in chemical models.

The low apparent CO to dust ratio observed in all disks has in general been attributed
to the freeze-out of CO onto the dust grains in the outer disk regions at
$r \ga 200$~AU with $T\la 20$~K. The importance of this mechanism is demonstrated by
the observations of CO isotopologues in the HD~163296 disk by \citet{2013arXiv1307.3420M}.
However, this mechanism, although unavoidable, may not be sufficient to explain the low CO content.
Strong apparent CO depletion (factor $\sim 100$) has also been observed in warm disks, with temperatures
above 30 K, where thermal desorption should occur: the disks around the Herbig stars CQ~Tau and MWC~758,
\citet{Chapillon_ea08}, and BP Tau \citep{Dutrey_ea03}.
\citet{Chapillon_ea08} suggested that CO may have been removed from the gas phase during
the cold pre-stellar phase by adsorption on small grains. During the warmer protostellar
phase, grain growth occurs, and CO may stay locked in the ice mantles of large grains that may
remain sufficiently colder than small grains during the reheating phase.
Also, from a complete modelling of CO isotopologues in TW Hya, \citet{Favre_ea13} concluded that
depletion alone could not account for the low CO to H$_2$ ratio, the H$_2$ content being derived
from detection of HD \citep{2013Natur.493..644B}. They invoke CO conversion to carbon chains, or perhaps
CO$_2$, that can remain locked in ice mantles at higher temperature than CO.
\begin{figure}
 \epsscale{1.0}
 \plotone{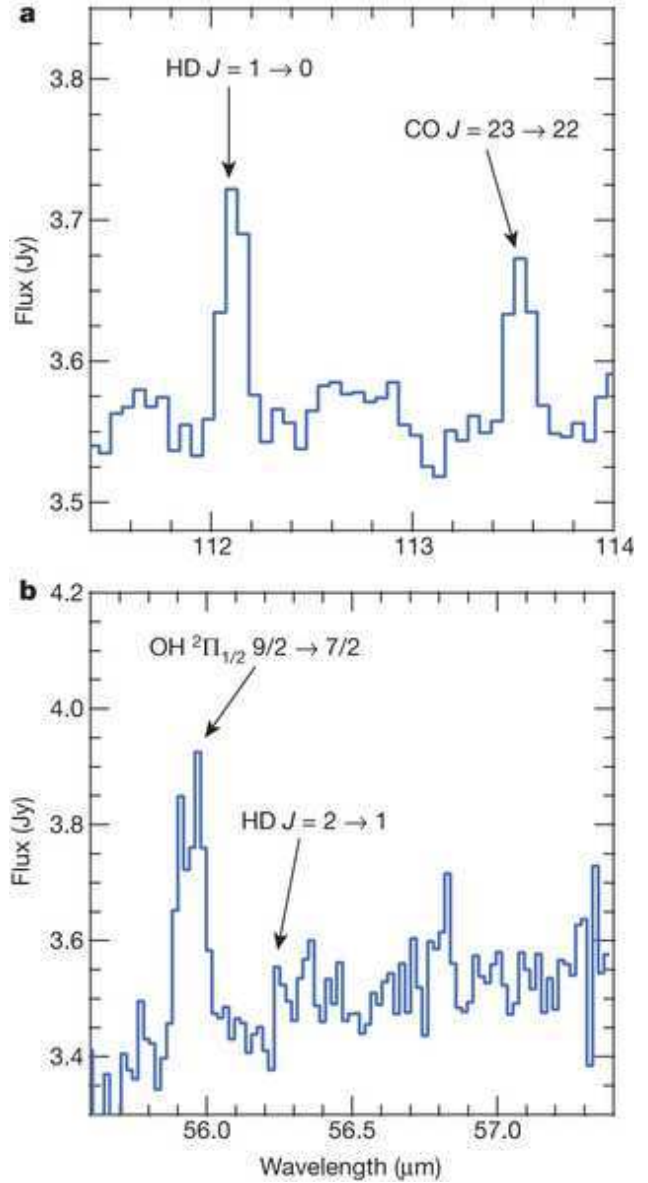}
 \caption{From \cite{2013Natur.493..644B}. Hydrogen Deuteride around TW Hydra observed with Herschel. The HD J=1--0 line at
112$\mu$m (a)
 is detected at 5$\sigma$ level, while there is only an upper limit on the HD J=2--1 line (b).}
 \label{spectra}
\end{figure}

\paragraph{Molecular complexity:}
Prior to ALMA, with the exception of HC$_3$N \citep{Chapillon_ea12b}, the molecules which have been
detected are the simplest, lighter molecules. \citet{2013ApJ...765L..14Q} 
recently report the
detection with ALMA of c-C$_3$H$_2$ in the warm disk surrounding the Herbig Ae star HD~163296 (see Fig.\ref{hd163296}).

\paragraph{Perspectives:} A major challenge of ALMA will be to refine our knowledge on
the vertical and radial structure by studying the excitation conditions
through multi-isotope, multi-transition studies of several molecular tracers
such as CN, CS, {\dbf HCO$^+$ or HCN at a high angular resolution. Particularly,
the CO isotopologues will remain robust tracers of the disk structure}.
Defining the physical conditions in the midplane will likely remain difficult and
should require long integration time to detect species such as H$_2$D$^+$,
DCO$^+$ or N$_2$D$^+$.

\bigskip
\noindent
\textbf{3.3 Inner Disk Structure ($<$30 AU)}
\bigskip

Unlike the outer disk, the inner ($<$30 AU) disk
has remained unresolved with (sub-)millimeter interferometers
\citep[see e.g.][]{Dullemond_Monnier2010}. In the near future,
the longest baselines of ALMA will start to image these regions, but
up to now all information about molecular gas has been obtained from
{\it spatially} unresolved observations. However, {\it spectrally}
resolved observations can be used to establish the region from which
line emission originates if a Keplerian velocity curve is assumed.
Infrared interferometry provides information on the innermost regions
at radii of $<1$ AU \citep[e.g.][]{Kraus2009}, but has been mostly
limited to continuum observations, with the exception of several solid-state
features \citep[e.g.][]{2004Natur.432..479V}.

SED modeling also provides at first order a good description of the density and
temperature structure of the dust throughout the disk.
However, especially in the inner disk and also at several scale heights,
gas temperatures may exceed  dust temperatures because of high-energy radiation
\citep{2004ApJ...615..991K}.
The (local) gas-to-dust ratio, dust-size distribution and
the effects of dust settling are all factors in determining the
relative dust and gas temperatures, and complicate the interpretation
of molecular line observations of the inner disk. Of particular
interest for the inner disk are constraints on any gaps present in
this region \citep[e.g.][]{2013A&A...555A..64M} and the shape of the gap
walls \citep{McClure2013} that depend on the gap-opening mechanism and
accretion \citep{2013arXiv1306.4264M}. {\dbf The disk gaps and inner
holes} provide a directly irradiated, warm surface visible in excited
molecular lines, and the question whether any gas remains in these regions
(see Chapters by Espaillat et al. and Pontoppidan et al.).

In recent years, the HIFI and PACS instruments on the Herschel Space
Observatory have probed the inner disks through spatially unresolved,
but, in the case of HIFI, spectrally resolved, emission lines of gas
species and several solid-state features from the dust. Several
authors \citep{2012A&A...544A..78M,2012A&A...538L...3R,2013A&A...555A..67R,2012A&A...544L...9F}
present Herschel observations of H$_2$O, CO, [O~I], OH, CH$^+$ and
[C~II] lines from T Tauri and Herbig Ae/Be disks. Atomic oxygen is
firmly detected in most disks and correlates with the far-IR dust
continuum, CO is observed in $\sim$ 50\% of disks and is stronger in
flaring disks, while [C~II] emission is often not confined to the disk
and difficult to separate from the surrounding environment
\citep{2012A&A...544L...9F,fedele2013}. H$_2$O and OH lines are only
observed in a few Herbig
stars \citep{2012A&A...544A..78M,2012A&A...544L...9F}, while they are
strong in T~Tauri stars with outflows \citep{2012A&A...545A..44P}.
Furthermore, the emission of hot water located around 2--3 AU is
observed in 24\% of gas-rich T Tauri
disks \citep{2012A&A...538L...3R}. Lastly, CH$^+$, tracing hot gas
was discovered in two Herbig stars with high UV flux: HD~97048 and
HD~100546 \citep{2011A&A...530L...2T,2012A&A...544A..78M}.

Using thermo-chemical models with a disk structure derived from
continuum observations, {\dbf it was shown that }these lines probe the inner disk as well as
the upper disk layers directly illuminated, and heated, by ultraviolet
radiation, where PAH heating can play an important
role \citep[e.g.][]{2010MNRAS.405L..26W,Bruderer_ea12}. Herbig
stars and the T~Tauri stars have an important difference in this
context: while in Herbig stars the UV radiation is stellar, in T~Tauri
stars it is mainly due to accretion {\dbf shocks}. Almost
universally, these lines show that the gas in the upper disk regions
is  warmer than the
dust \citep[e.g.][]{2004ApJ...615..991K,Bruderer_ea12}, in
accordance with models including photon heating (PDRs). Considering
multiple transitions of CO in the HD~100546 disk, \citet{fedele2013}
found a steeper temperature gradient for the gas compared to the dust,
providing further direct proof of thermal decoupling between gas and
dust. \citet{2012A&A...547A..69A} showed that the [O~I] line flux
increases with UV flux when L$_X < 10^{30}$ erg s$^{-1}$, and with
increasing L$_X$ {\dbf (when it is higher that the UV luminosity)}. \citet{2011A&A...530L...2T} detected
CH$^+$ in HD~100546, and concluded that this species is most abundant at
the disk rim at 10--13 AU. In HD~163296,
\citet{2012A&A...538A..20T}
studied the effect of dust settling, and found that in settled models
the line fluxes of species formed deeper in the disk are
increased. \citet{2013arXiv1304.5718T} found that the line observations
of the disk around 51~Oph can only be explained if is compact
($<$10--15 AU), although an outer, cold disk devoid of gas {\dbf may be extended up}
to 400 AU. For all these lines observed with Herschel, it is good to
remember that, while they dominate the far-infrared spectrum, they
only trace a small fraction of the disk, and that the dominant disk
mass reservoir near the midplane at these radii is much cooler and
will require high-resolution millimeter (ALMA) observations.

Solid-state features present in the Herschel wavelength range provided
information about the dust composition of the inner disk.
{\bbf Emission of the 69 $\mu$m forsterite feature in the disk of HD~100546
\citep{2010A&A...518L.129S,2013A&A...553A...5S} was shown by
\citet{2011A&A...531A..93M} to be dominated by
emission from the inner disk wall located between 13--20 AU.}
Detailed modeling of the line shape of the feature yielded a crystalline mass
fraction of 40--60\% in this region, with a low iron content of
$<$0.3\%.{\bbf A tentative detection of crystalline water ice using PACS
was also reported by \citet{McClure+2012} in the disk surrounding the T~Tauri star GQ Lup.}

The chapter of Pontoppidan  {\dbf et al.} in this volume extensively
discusses the volatile content of disks. Here we only mention that
significant {\dbf decrease of the H$_2$O abundances} across the expected snowline has been
observed in several disks using unresolved Herschel and Spitzer
observations \citep[e.g.][]{2009ApJ...704.1471M,2013ApJ...766...82Z}.
\citet{2013ApJ...766..134N} and \citet{2013ApJ...765...34Q} concluded that species such HCN, N$_2$H$^+$, and H$_2$CO
{\dbf reveal} the
presence of H$_2$O and CO
snowlines through chemical signatures.

Intriguing conclusions can be drawn from spatially unresolved
observations using spectrally resolved emission lines, when an
underlying Keplerian velocity profile is assumed. Using Science
Verification data from ALMA {\dbf and the high signal-to-noise CO line data},
\citet{2012ApJ...757..129R} found that the inner disk of TW Hya is likely warped by a few degrees on scales of 5
AU. Such a warp is
consistent with what was earlier deduced by optical scattered light
\citep{2005ApJ...622.1171R}, and may be explained by a (planetary?)
companion inside the disk.

With the roll-out of ALMA, high sensitivity imaging of
regions as small as a few AU will become feasible,
providing direct information on the region where planets form.

\bigskip
\centerline{\textbf{4. THERMO-CHEMICAL PROCESSES}}
\bigskip
\begin{figure*}
 \epsscale{1.5}
\center
\includegraphics[angle=90.0,width=17.0cm]{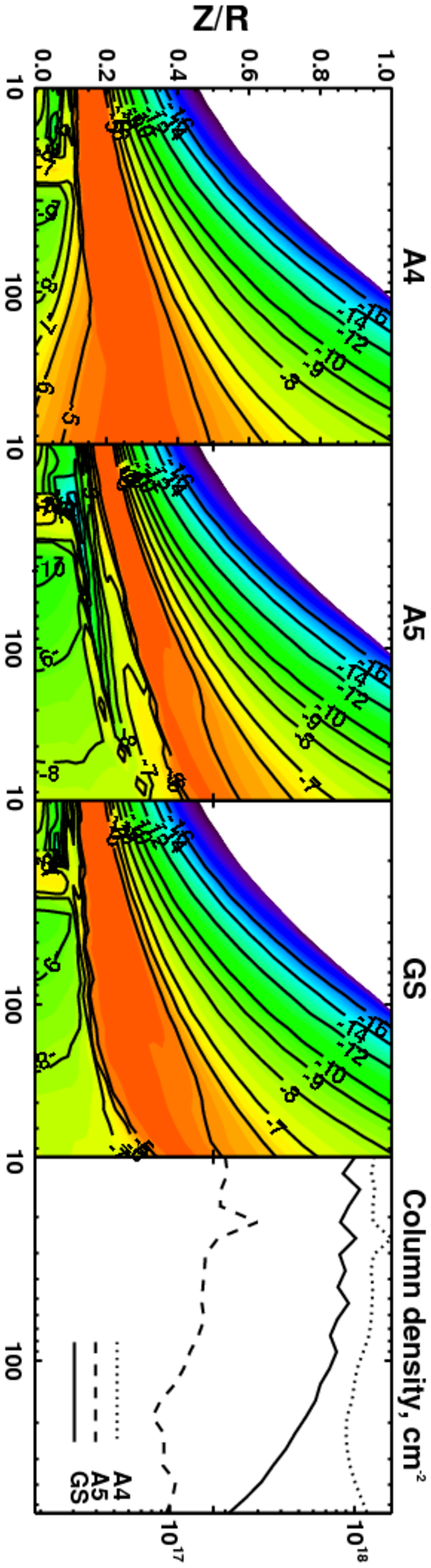}
\includegraphics[angle=0.0,width=11.0cm]{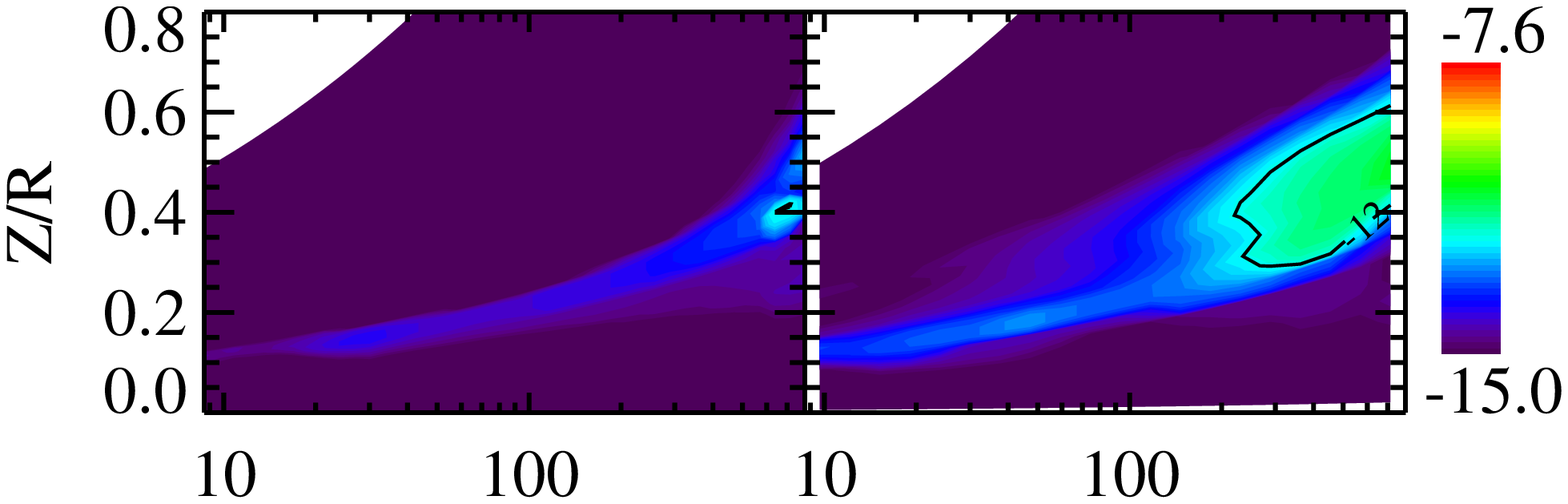}
\includegraphics[angle=0.0,width=11.0cm]{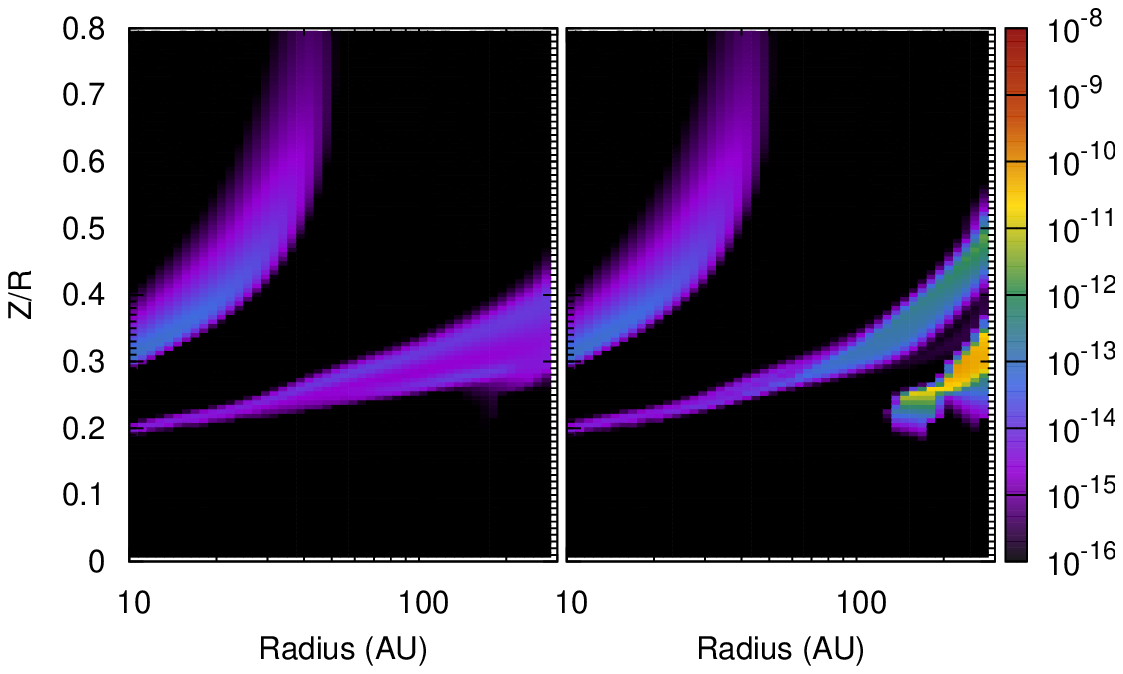}
\caption{\dbf Top panel: Fractional abundance profiles (first three columns) and column densities (fourth column) of
the gas-phase carbon monoxide in a T Tauri disk calculated with three distinct dust grain models:
(A4) the uniform grains with radius of $1~\mu$m, (A5) the uniform grains with radius of 0.1 $\mu$m
and (GS) the model with grain growth and settling.
Note how grain evolution affects the depletion of molecules onto grains
near the midplane of cold outer disk (\cite{Vasyunin2011}).
Middle panel: Gas-phase methanol abundances in a T Tauri disk calculated with
a laminar gas-grain chemical model (left) and a 2D turbulent mixing
chemical model (right). The transport of methanol ice from dark midplane to
upper, more UV-irradiated disk layer enables more methanol ice to be desorbed in the gas
phase (\cite{Semenov_Wiebe11a}).
Bottom panel: Gas-phase methanol abundances in a T Tauri disk calculated with
two gas-grain chemical models: (left) no surface chemistry is taken account
and (right) with surface chemistry (\cite{Walsh_ea10}). The surface hydrogenation of CO
leads to efficient production of methanol ice, which can be partly UV-desorbed
in disk layers above midplane.
}
\label{model}
\end{figure*}

Protoplanetary disks are characterized by strong vertical and radial temperature and
density gradients and varying UV/X-ray radiation intensities. These conditions favors
 diverse  chemical processes, including photochemistry, molecular-ion reactions, neutral-neutral reactions,
gas-grain surface interactions, and grain surface reactions. Local
density, temperature and high-energy irradiation
determine which of these processes will be dominating.

Disk chemistry is driven by high-energy radiation and cosmic
rays \citep{ah1999,Aea02,vZea03,Gorti_Hollenbach04,2006PNAS..10312249V,vDea_06,Fogel_ea11,Vasyunin2011,Walsh_ea12,ANDES}.
T~Tauri stars possess intense non-thermal UV radiation from the accretion shocks,
while the hotter Herbig~Ae/Be stars produce intense thermal UV emission. The overall intensity of the stellar UV
radiation at 100~AU can be higher by a factor of
$100-1\,000$ for a T~Tauri disk \citep{Bergin_ea04} and 10$^5$ for a Herbig~Ae disk \citep{Semenov_ea05,Jonkheid_ea06},
respectively, compared to the  interstellar radiation field \citep{G}.
Photodissociation of molecules will depend sensitively on the shape and strength of the UV radiation field.
For example, Lyman $\alpha$ photons will
selectively dissociate HCN and H$_2$O, while other molecules that dissociate between 91.2 and 110~nm such as CO and H$_2$ remain
unaffected \citep{1988ApJ...334..771V,vDea_06}. Selective photodissociation of CO by the interstellar
radiation field can also play an important role in the outer disk region \citep{DDG03}.

The $\sim 1-10$~keV X-ray radiation is another important energy source for disk chemistry.
The  median value for the X-ray luminosity of the T~Tauri
stars is $\log (L_{\rm X}/L_{\rm bol}) \approx -3.5$ or $L_{\rm X} \approx 3\,10^{29}$~erg\,s$^{-1}$
(with an uncertainty of an order of
magnitude \citep[see e.g.][]{Preibisch_ea05,Getman_ea09}. This radiation is generated by coronal activity,
driven by magnetic fields generated by an $\alpha\omega$ dynamo mechanism
in convective stellar interiors. Herbig~Ae stars have weak surface magnetic fields due
to their non-convective interiors. Their typical X-ray
luminosities are $\ga 10$ times lower than those of T~Tauri stars \citep{Guedel_Naze09}.
The key role of X-rays in disk chemistry is their ability to penetrate through a high gas columns ($\sim 0.1-1$~g\,cm$^{-2}$)
and ionize He, which destroys tightly bound molecules like CO, and replenishes gas with elemental carbon and oxygen.
The similar effect for disk chemistry is provided by cosmic-ray particules (CRPs), which can penetrate even through very high gas columns,
$\sim 100$~g\,cm$^{-2}$ {\dbf \citep{un1980,1996ApJ...457..355G,2013ApJ...772....5C}.}

In the terrestrial planet-forming region, $\sim 1-10$~AU, temperatures are $\ga 100$~K and densities may exceed
$10^{12}$~cm$^{-3}$.
Despite intense gas-grain interactions, surface processes do not play a role and disk chemistry
is determined by fast gas-phase reactions, which reaches quickly a steady-state.
In the absence of intense sources of ionizing radiation and high temperatures  neutral-neutral reactions with
barriers become important \citep{Harada_ea10}. At the densities  $\ga 10^{12}$~cm$^{-3}$ 3-body reactions also become
important \citep{Aea99}.

The outer disk regions ($ \ga 10$~AU) can be further divided into 3 chemically different regimes \citep{ah1999}:
(1) cold chemistry in the midplane dominated by surface processes, (2) rich gas-grain chemistry in the
warm intermediate layer, and (3) restricted PDR-like gas-phase chemistry in the atmosphere.

In the disk atmosphere stellar UV radiation and the interstellar radiation field ionize
and dissociate molecules and drive gas-phase PDR-like  chemistry.
Adjacent to the disk surface a warm, $\approx
30-300$~K, molecular layer  is located. This region is partly shielded from stellar and interstellar UV/X-ray radiation,
and chemistry is active in the gas and on the dust surfaces.
The ionization produces ions, like H$_3^+$, driving rapid proton transfer and other ion-molecule
processes \citep{HerbstKlemperer73,Watson_74a,Woodall_ea07,KIDA}.
If water is frozen onto dust grains, a relatively high C/O
ratio close to or even larger than 1 can be reached in the gas phase,
leading to a carbon-based chemistry. Ices are released from grains
by thermal or photodesorption.

In the outer midplane, the cosmic ray particles and locally produced UV photons are the only ionizing sources.
The temperature drops below $\sim 20 - 50$~K and freeze-out of molecules and hydrogenation reactions on grain surfaces are
dominating the chemistry.
The most important desorption process for
volatiles such as CO, N$_2$, and CH$_4$ is thermal desorption. In addition, cosmic ray and X-ray spot heating may
release mantle material back to the gas phase \citep{Leger_ea85,HH93,Najita_ea01}, {\dbf as well as reactive
desorption \citep{Garrod_ea07}}.
In the outer, cold disk regions addition of nuclear-spin-dependent chemical reactions
involving ortho- and para-states of key species is required.

\bigskip
\noindent
\textbf{ 4.1 Line Emission and Thermo-Chemical Models}
\bigskip

Increasing detections of molecular, atomic and ionic emission lines from
ground-based infrared and (sub)mm facilities and  {\em
Spitzer} and {\em Herschel} space observatories have led to a recent
shift in emphasis on modeling the {\em gas} structure in disks.

Gas line emission
is sensitive to both the abundances of trace emitting species (and hence
chemistry) and the excitation conditions (density, temperature,
irradiation) at the surface. Meanwhile, the dominant cooling process
which controls the gas temperature in the disk surface is radiative
cooling by transition lines (e.g. OI, CII, Ly $\alpha$). Therefore, we
need to treat chemistry and
thermal processes self-consistently in order to obtain gas temperature
structure and predict line fluxes. Thermo-chemical models are important
also in order to understand photoevaporation process as a gas dispersal
mechanism from disks {\dbf \citep[][see also Chapter by Espaillat et al.]{2009ApJ...690.1539G,Gorti_ea09}.}

\bigskip
\noindent
\textit{4.1.1 Thermo-Chemical Models}
\bigskip

Early disk models assumed that the gas temperature was equal to that of
the dust. In the surface layers, however, low densities and/or  heating
of gas by stellar X-ray/UV radiation result in weaker thermal coupling
of gas to dust and gas and dust temperatures can significantly
differ. This is more so in disks where the dust has evolved and the
collisional cross-sections are lower (see Sect.4.2.1). Importantly,
the vertical density distribution is determined by the gas thermal
pressure gradient. These considerations have given rise to the
development of thermo-chemical disk models where the gas and dust
temperatures are determined separately and self-consistently solved with
chemistry to allow more accurate interpretations of observed line
emission \citep[see also ][]{2007prpl.conf..555D}.

The first thermo-chemical models consistently solved for the gas
temperature structure coupled with chemistry, using a density
distribution derived from the dust radiative transfer and temperature
solutions. Temperature of gas deviated significantly from dust in
optically thin dust-rich disks \citep{Kamp_Zadelhoff01}.  These
models were extended to optically thick disks \citep{2004ApJ...615..991K}
and it was determined that gas/dust temperatures were well-coupled in
the midplane but not in the higher optically thin dust surface. UV
heating was found to be important in the thermally decoupled regions,
probed by molecules such as H$_2$O, CO and OH.
\citet{Jonkheid_ea04,Jonkheid_ea07} considered the effects of dust
evolution to find that dust settling and the resulting increased
decoupling allowed gas temperatures to rise; this often increased the
intensities of atomic and molecular lines.
While these studies focused on UV heating of gas, Glassgold and
collaborators \citep[e.g.][]{Glassgold_ea04,Glassgold_ea07,Glassgold_ea09,
2012ApJ...756..157G, Meijerink_ea08a,2011ApJ...743..147N}
have been
examining the effect of X-rays on gas heating and chemistry with
increasing levels of detail. X-ray heating dominates surface gas heating
in these models and leads to ionization deep in the disk, driving
ion-molecule chemistry and the formation of water and other organics.
\citet{Glassgold_ea07} predicted an important strongly emitting tracer of ionized gas, the mid-infrared line of [NeII] at
12.8$\mu$m. The primary conclusion of these investigations is that disk structure  closely resembles photo-dissociation regions
(PDRs and XDRs), with a cold, shielded molecular interior and warmer, atomic/ionic surface regions.

A further refinement to thermo-chemical models was to self-consistently
solve for vertical hydrostatic equilibrium in the disk as set by the gas
temperature structure \citep{Gorti_Hollenbach04,Gorti_Hollenbach08,
Nomura_Millar05,Aikawa_ea06,Nomura_ea07a} and to consider
irradiation of the disk surface by both UV and X-rays
\citep{Gorti_Hollenbach04,Gorti_Hollenbach08,Nomura_ea07a,ANDES}.
Gas in the surface layer is hotter than the dust grains, which
results in a more vertically extended disk atmosphere.

The increased UV/X-ray attenuation in the inner disk lowers the surface
gas temperature at intermediate and larger radii compared to the
solutions obtained without a gas-determined equilibrium density
structure. X-ray heating is found to dominate in the surface layers of
disks, often heating gas to $\gtrsim$ 1000-5000K in the inner disk,
while FUV heating and both X-ray and UV photodissociation are important
in the intermediate (A$_{\rm V} \sim 0.1-1$) regions. Dust thermal
balance dominates in the dense, shielded interior layers, although
ionization by hard X-ray photons can drive ion-molecule chemistry even
in the denser regions of the disk.  The change in temperature obtained
with a self-consistent density determination is inferred to affect the
calculated intensities of molecular and atomic line emission from
disks.

More recently \citet{ANDES} have introduced a state-of-the-art
thermo-chemical ANDES model, which treats time-dependent chemical
reactions self-consistently with {\dbf gas and dust} thermal processes and
calculations of dust coagulation, fragmentation and settling
(see Sect.4.2.1).


\bigskip
\noindent
\textit{4.1.2 Modeling Line Emission}
\bigskip

Recent development of thermo-chemical models focused more on {\dbf modeling
observable line emissions} from disks.
\citet{Gorti_Hollenbach08}
found that [NeII] and [ArII] lines are tracers of X-ray ionized regions, and that
X-rays and FUV are both significant contributors to gas heating. They concluded that
[OI] 63$\mu$m line was a strong, luminous coolant in disks.
\citet{Nomura_ea07a} investigated the strength of various H$_2$ lines,
including dust coagulation calculations, to suggest that infrared
line ratios of ro-vibrational transitions of H$_2$ are sensitive to gas
temperature and FUV radiation field, and could be useful tracer of grain
evolution in the surface layers of the inner disks.

In order to statistically model and predict line fluxes from disks for
observations by Herschel, ALMA, and forthcoming facilities, a tool,
called the DENT grid, was developed to
calculate line fluxes with a large number of parameters, such as stellar
mass, age, and UV excess, disk masses, scale heights,
inner and outer radii, dust properties, and inclination angles,
as a part of the Herschel open-time key program of GASPS
\citep{2010MNRAS.405L..26W,Kamp_ea11a}. The DENT grid used the
3D Monte Carlo radiative transfer code, MCFOST \citep{Pinte_ea06,
2009A&A...498..967P} for calculating dust temperature and line radiative
transfer, and the thermo-chemical model, ProDiMo, for calculating gas
temperature and abundances of species.
%
ProDiMo is a sophisticated model which takes into consideration all
of the processes mentioned in the
previous subsection. \citet{Woitke_ea09} utilized full 2-D dust continuum
transfer, a modest chemical network, and {\dbf computed} the gas temperature
and density structure (in 1+1D), but initially only including UV heating and
{\dbf simplified photo}chemistry. These were updated to improve upon the calculations of the
rates of photoprocesses and line radiative transfer
\citep{2010A&A...510A..18K}. X-ray irradiation was added in
\citet{Aresu_ea10a} and \citet{Meijerink_ea12a}.
\citet{2011MNRAS.412..711T} studied the effects of the inner rim structure on the disk.
These models have been successfully applied to observational data
of Herschel GASPS objects of protoplanetary disks and debris disks
to infer the important disk properties, such as dust evolution,
radial extent, and disk masses \citep{2010A&A...518L.124M,
Thi_ea11a,2012A&A...538A..20T,2013ApJ...766L...5P}.

\citet{Bruderer_ea09a} {\dbf developed another thermo-chemical disk model.
\cite{Bruderer_ea12} adapted it to a disk around a Herbig Be
star, HD~100546, and used it to explain the high-$J$ CO lines and the [OI] 63$\mu$m
line observed by Herschel, together with an upper limit on the [CI] 370
$\mu$m line (by APEX). They discussed the variability of} the gas/dust ratio and the amount of
volatile carbon in the disk atmosphere.
\citet{Chapillon_ea08} adopted the Meudon PDR code
\citep{1993A&A...267..233L,2006ApJS..164..506L} with a number of
{\dbf variable} parameters such as UV field, grain size, gas-to-dust ratio to explain
millimeter CO line observations towards Herbig Ae disks.

While the thermo-chemical models have been developed, the non-LTE line
radiative transfer methods for calculating transition lines from disks
also have been studied.
In \citet{Pavlyuchenkov_ea07a} the comparison between various approximate line radiative transfer
approaches and a well-tested accelerated Monte Carlo code was performed. Using a T~Tauri-like
flared disk model and various distributions of molecular abundances (layered and homogeneous),
the excitation temperatures, synthetic spectra, and channel maps for a number of rotational
lines of CO, C$^{18}$O, HCO$^+$, DCO$^+$, HCN, CS, and H$_2$CO were simulated. It was found that
the LTE approach widely assumed by observers is accurate enough for modeling the low molecular
rotational lines, whereas it may significantly overestimate the high-$J$ line intensities
in disks. The full escape probability (FEP) approach works better for these high-$J$ rotational
lines but fails sometimes for the low-$J$ transitions.  \citet{Semenov_ea08} adopted the code
to simulate the ALMA observations of maps of molecular line emission from disks, using the resulting
molecular abundance profiles of their calculations of chemical reactions.

\citet{2010A&A...523A..25B} developed a line radiative transfer code, LIME,
based on the RATRAN code \citep{RATRAN}, in which grids are laid out
adaptive to the opacity. Therefore, the code can deal with objects having
inhomogeneous 3D structure with large density contrast as well as
overlapping lines of multiple species.
The code is available on the website: http://www.nbi.dk/$\thicksim$brinch/lime.php.
{\dbf The LIME code is used in the ARTIST software package designed to model 2D/3D
line and continuum radiative transfer and synthesize interferometric images
\citep[][http://youngstars.nbi.dk/artist/Welcome.html]{2012A&A...543A..16P}.}

Non-LTE effects are more significant for higher transition lines whose
critical densities are high. \citet{2009ApJ...704.1471M} modeled mid-infrared
water lines observed by Spitzer 
\citep{HITRAN} and showed that the differences in line fluxes obtained by LTE and
non-LTE calculations are within an order of magnitude for most of the
lines. \citep{2009A&A...501L...5W} also studied water line {\dbf fluxes,
including the submillimeter  lines and using their full
thermo-chemical model. They showed that the differences are smaller for low
excitation water} lines.

\bigskip
\noindent
\textbf{4.2 Chemistry vs. Physical Processes}
\bigskip

Dynamical processes which lead to planet formation, such as grain
evolution and gas dispersal, affect chemical structure in protoplanetary disks.
Thus, these processes as well as some environmental effects could appear
in molecular line emission from the disks.

\bigskip
\noindent
\textit{4.2.1 Effect of Grain Evolution}
\bigskip

Grain size distributions in protoplanetary disks are thought to be very
different from that of the interstellar grains because of coagulation,
settling towards the disk midplane and transport towards the central
star under the influence of the gravity of the star (see chapter
by Testi et al.).

One of the recent advancement in studies of disk chemistry is the
treatment of grain evolution.
Grain growth
depletes the upper disk layers of small grains and hence reduces the opacity of disk matter, allowing the far-UV radiation
to penetrate more efficiently into the disk, and to heat and dissociate molecules deeper in the disk.
Also, larger grains populating the disk midplane will delay the depletion of gaseous
species because of the reduced total surface area {\dbf \citep{Aikawa_ea06,Fogel_ea11,Vasyunin2011}}.
Grain coagulation, fragmentation,
sedimentation, turbulent stirring and radial transport are all important
processes to control the grain size distribution.

\citet{Nomura_ea07a} investigated the effect of dust evolution only on gas
temperature profile and molecular hydrogen excitation by using the
result of calculations of coagulation equation with settling of dust
particles.
\citet{Vasyunin2011} studied the effect on chemical reactions, taking
into account the fragmentation and cratering of dust particles in
addition.

{\bbf \citet{ANDES} have developed the ANDES model,} which is based on combined description of the 1+1D continuum
radiative transfer, detailed 2D dust evolution model,
and time-dependent laminar chemistry. This model was applied to investigate
the impact of grain evolution on disk chemistry. It was
found that due to grain growth and dust removal from the disk atmosphere,
the molecular layer shifts closer toward the disk
midplane that remains shielded by dust from the ionizing radiation. Larger
grains settled to the midplane lower the frequency of  gas-grain
collisions, and thus making depletion of gaseous molecules
less severe and
hindering the growth of ices on grains.

\bigskip
\noindent
\textit{4.2.2 Chemo-Dynamical Models}
\bigskip

Gas in protoplanetary disks is thought to disperse in $1-10$ Myr.
Magneto-rotational instability driven turbulence will cause angular
momentum transfer and the gas accretes towards the central star as a
result, while photoevaporation disperses the gas in the region where
the thermal energy is high enough for the gas to escape from the gravity
of the central star (see chapters by Turner et al., Alexander et al. and
Espaillat et al.).
The gas motion associated with this gas dispersal process is expected to
affect the chemical structure, especially near the boundaries between
the three layers: photon-dominated surface layer, warm molecular-rich
middle layer, and cold molecule-depleted layer near the midplane.
The effect appears significant when the timescale of the gas motion is
shorter than the timescales of the dominant chemical reactions. So far,
in addition to laminar disk models, a number of chemo-dynamical models
of protoplanetary disks have been developed.

On the other hand, chemo-dynamical models have been also developed to treat
chemistry in dynamically evolving early phase of star and disk formation
where circumstellar disks are still embedded in dense envelope.
Some information in this early phase could remain of chemistry in
protoplanetary disks after the surrounding envelope clears up,
especially of the ice molecules on grains.

In the coming years, we expect that sophisticated multi-dimensional magneto-hydrodynamical models will be coupled with
time-dependent chemistry. First steps in this direction have been made \citep{Turner_ea07}, where a simple chemical model has
been coupled to a local 3D MHD simulation.

\bigskip
\noindent
\textit{4.2.2.1 Turbulent Mixing and Gas Accretion}
\bigskip

Chemical evolution in protoplanetary disks with 1D radial advective mass
transport was studied by
\citet{Woods_Willacy07,Woods_Willacy08,Willacy_Woods09,Nomura_ea09}.
Another class of chemo-dynamical disk models is based on turbulent diffusive
mixing, which uniformises chemical gradients, and is modeled in 1D, 2D, or
even full 3D
\citep{Ilgner_Nelson06,Ilgner_Nelson06a,Ilgner_Nelson06b,Ilgner_ea08,Semenov_ea06,Aikawa_07,Semenov_Wiebe11a,Willacy_ea06,
Hersant_ea09,Turner_ea06,Turner_ea07}. In several disk models both advective and
turbulent transport was considered \citep{Heinzeller_ea11}, while \cite{Gorti_ea09} studied photoevaporation of disks
and loss of the gas due to the stellar far-UV and X-ray radiation.

The turbulence in disks is a 3D-phenomenon driven by
a magneto-rotational instability \citep[][,see also Turner et al. chapter in this book]{MRI}.
Global MHD simulations show that advection has no specified
direction in various disk regions, and in each location goes both inward and outward.
The corresponding turbulent velocity of the gas $V_{\rm turb}$ depends on the viscosity parameter {\bbf $\alpha$
\citep{1973A&A....24..337S}}
and scales with it somewhere between linear and square-root dependence: $\alpha  <  V_{\rm turb}/c_{\rm s} < \sqrt{\alpha}$,
here $c_{\rm s}$ is the sound speed.
The calculated $\alpha$-viscosity stresses have values in a range of $10^{-4}$ and $10^{-2}$ in the midplane and the molecular
layer,
respectively, and rise steeply to transonic values of $\sim 0.5$ in the disk atmosphere.

MRI requires a modest gas ionization to be operational.
The ionization structure of protoplanetary disks is largely controlled by a variety of chemical processes.
Recent studies of the chemistry coupled to the dynamics in protoplanetary disks are briefly summarized below
(see also Sect. 5.1). \citet{Ilgner_Nelson06b} have studied the
ionization structure of inner disks ($r<10$~AU), considering
vertical mixing and other effects like X-ray flares, various elemental compositions, etc.
They found that mixing has no profound effect on electron concentration if metals are absent in the gas since recombination
timescales are faster than dynamical timescales. However, when $T\ga 200$~K and alkali atoms are present in the gas,
chemistry of ionization becomes sensitive to transport, such that diffusion may reduce the
size of the turbulent-inactive disk ``dead'' zone (see also Sect. 5.2).

\citet{Willacy_ea06} have attempted to study systematically the impact of disk diffusivity on
evolution of various
chemical families. They used a steady-state $\alpha$-disk model similar to that of \citet{Ilgner_ea04} and considered
1D-vertical mixing in the outer disk region with $r>100$~AU. They found that vertical transport can increase column
densities (vertically integrated concentrations) by up to 2 orders of magnitude for some complex species.
Still, the layered disk structure was largely preserved even in the presence of vertical mixing.
\citet{Semenov_ea06} and \citet{Aikawa_07} have found that turbulence allows to
transport gaseous CO
from the molecular layer down towards the cold midplane where it otherwise remains frozen out, which
may explain the large amount of cold ($\la15$~K) CO gas detected in the disk of DM~Tau \citep{DDG03}.
\citet{Hersant_ea09} have studied various mechanisms to retain gas-phase CO in very cold disk regions,
including vertical mixing. They found that  photodesorption in upper, less obscured molecular layer greatly increases
the gas-phase CO concentration, whereas the role of vertical mixing is less important.

Later, in \citet{Woods_Willacy07} the formation and destruction of benzene in turbulent protoplanetary
disks at $r\la 35$~AU has been investigated. These authors found that radial transport allows efficient synthesis of benzene at
$\la 3$~AU, mostly due to ion-molecule reactions between C$_3$H$_3$ and
C$_3$H$_4^+$ followed by grain dissociative recombination. The resulting concentration of C$_6$H$_6$ at
larger radii of $10-30$~AU is increased by  turbulent diffusion up to 2 orders of magnitude.

In a similar study, \citet{Nomura_ea09} have considered a inner disk model with radial advection
($\la 12$~AU). They found that the molecular concentrations are sensitive to the transport speed, such that
in some cases gaseous molecules are able to reach the outer, cooler disk regions where they should be depleted.
This increases the production of many complex or surface-produced species such as methanol, ammonia,
hydrogen sulfide, acetylene, etc.

\citet{TG07} have considered
a 2D disk chemo-hydrodynamical model in which global circulation flow patterns exist, transporting disk matter
outward in the disk midplane and inward in elevated disk layers. They found that
gas-phase species
produced by warm neutral-neutral chemistry in the inner region can be  transported into the cold outer region and
freeze out onto dust grain surfaces. The presence of such large-scale meridional flows in protoplanetary accretion disks
was called in question in later MHD studies \citet[][]{2011A&A...534A.107F,2011ApJ...735..122F}.

\citet{Heinzeller_ea11} have studied the chemical evolution of a protoplanetary disk with
radial viscous accretion, vertical mixing, and a vertical disk wind (in the atmosphere). They used a steady-state disk
model with $\alpha=0.01$ and $\dot{M}= 10^{-8}\,M_\odot$\,yr$^{-1}$. They found that mixing lowers concentration
gradients, enhancing abundances of NH$_3$, CH$_3$OH, C$_2$H$_2$ and sulfur-containing species. They concluded
that such a disk wind has an effect similar to turbulent mixing on chemistry, while the radial accretion changes molecular abundances
in the midplane, and the vertical turbulent mixing enhances abundances in the intermediate molecular layer.

A detailed study of the effect of 2D radial-vertical mixing on gas-grain chemistry in a protoplanetary
disk has been performed in \citet{Semenov_Wiebe11a}. These authors used the $\alpha$-model of a
$\sim5$~Myr DM~Tau-like
disk coupled to the large-scale gas-grain chemical code ``ALCHEMIC'' \citep{Semenov_ea10}. To account for
production of complex molecules, an extended set of surface processes was added.
A constant value of the viscosity parameter $\alpha=0.01$ was assumed.

In this study it was shown that turbulent
transport changes the abundances of many gas-phase species and particularly ices.
Vertical mixing is more important than radial mixing, mainly because radial temperature and density gradients in disks
are weaker than vertical ones. The simple molecules not responding to dynamical transport include
C$_2$H, C$^+$, CH$_4$, CN, CO, HCN, HNC, H$_2$CO, OH, as well as water and ammonia ices.
The  species sensitive to transport are carbon chains and other heavy species,
in particular sulfur-bearing and complex organic molecules (COMs) frozen
onto the dust grains. Mixing transports ice-coated grains  into
warmer  disk regions where  efficient surface recombination reactions proceed.
In warm molecular layer these complex ices evaporate and return to the gas phase.

\bigskip
\noindent
\textit{4.2.2.2 Dynamics in Early Phase}
\bigskip

In protostellar phase, materials in envelopes surrounding circumstellar
disks dynamically fall onto the disks, and then accrete towards the
central star in the disks. Chemical evolution, including gas-grain interaction,
along 2D advection flow from envelopes to disks was studied by
\citet{Visser_ea09b,Visser_ea11a}. Using axisymmetric 2D semi-analytical dynamical models,
they discussed the origin of organic molecules and other species in comets in our Solar System.

Assumption of axisymmetry will be adoptable for quiescent disks where
mass is transported locally with low accretion rates, but
may be inappropriate for disks in the early phase, which
are massive enough to trigger
gravitational instabilities \citep{Vorobyov_Basu06}. Gravitational instabilities produce
transient non-axisymmetric structures in the form of spiral waves, density clumps, etc\ldots \citep{Boley_ea10a,Vorobyov_Basu10a}.
The results of Vorobyov and collaborators have been confronted by
\citet{Zhu2010a,Zhu2010b,Zhu2012},
who found that ongoing accretion of an envelope surrounding a forming
protoplanetary disks can alone drive non-stationary accretion and
hence outbursts similar to the FU~Ori phenomenon.
More importantly, gravitational instabilities lead to efficient
mass transport and angular momentum redistribution, which could be
characterized by a relatively high viscosity parameter $\alpha \sim 1$.
\citet{Ilee_ea11a} performed a first study of time-dependent chemistry
along paths of fluid elements in 3D radiative hydrodynamical simulations
of a gravitationally unstable massive disk ($0.39 M_{\odot}$ within a
disk radius of 50\,AU) by \citet{Boley_ea07} and showed importance of
continuous stirring by rotating spiral density waves with weak shocks on
chemical properties in the disk.

\bigskip
\noindent
\textit{4.2.3 Environmental Effects}
\bigskip

Since chemical processes are sensitive to local density, temperature,
and radiation fields, we expect to see different chemical structure and
molecular line emission if a protoplanetary disk is in a specific
environment.

\citet{Cleeves_ea11a} investigated the chemical structure of a
transition disk with a large central hole (with a radius of 45\,AU).
Transition disks are interesting objects to study dispersal
mechanisms of disks, for example, by giant planets and/or
photoevaporation, and so on (see chapters by Espaillat et al. and
Alexander et al., and Sect. 5.4). \citet{Cleeves_ea11a}
showed that the inner edge of the disk is heated by direct irradiation
from the central star so that the even
molecules with high desorption energies can survive in the gas-phase
near the midplane. Line radiative transfer calculations predicted
that this actively evolving truncation region in transition disks will
be probed by observations of high transition lines of such molecules
by ALMA with high spatial resolution.

\citet{2013ApJ...766L..23W} studied the chemical structure of a
protoplanetary disk irradiated externally by a nearby massive star.
Although it is observationally suggested that stars are often formed in
young star clusters \citep[e.g.][]{2003ARA&A..41...57L}, star and planet
formation in isolated systems have been mainly studied because detailed
observations have been available only isolated star forming regions
close to our Sun so far. However, ALMA is expected to provide detailed
(sub)millimeter observations of protoplanetary disks in young clusters,
such as the Trapezium cluster in the Orion nebula, allowing us to study star
and planet formation in varied environments. \citet{2013ApJ...766L..23W} showed
that external irradiation from nearby OB stars can heat gas and dust even
near the disk midplanes, especially in the outer regions,
affecting snowlines of molecules with desorption temperature of
$\sim$30K. Also, due to the externally irradiated FUV photons, the
ionization degree becomes higher in the disk surface. Predictions of
molecular line emission showed observations of line ratios
will be useful to probe physical and chemical properties of the disk,
such as ionization degree and photoevaporation condition.

\citet{2013ApJ...772....5C} investigated the ionization rate in protoplanetary
disks excluded by cosmic rays. It has been suggested that strong
winds from the central young star with strong magnetic activity may
decrease cosmic-ray flux penetrating into the disks by analogy with
the solar wind modulating cosmic-ray photons in our Solar System.
\citet{2013ApJ...772....5C}
showed that the ionization rate by cosmic-ray can be
reduced by stellar winds to a value lower than that due to short-lived
radionuclides of $\zeta_{\rm RN}\sim 10^{-18}$s$^{-1}$
\citep{2009ApJ...690...69U}, with which magnetorotational instability will be
locally stabilized in the inner disk.

\bigskip
\noindent
\textbf{4.3 New advances in treatment of chemical processes}

\bigskip
\noindent
\textit{4.3.1 Photochemistry}
\bigskip

Under the strong UV and X-ray irradiation from the central star,
photochemistry is dominant in the surface layers of protoplanetary
disks. A detailed 2D/3D treatment of the X-ray and UV radiation transfer
with anisotropic scattering is an essential ingredient for realistic
disk chemical models \citep[e.g.][]{IG99,vZea03}. The photochemistry
dominant region expands in the disks with grain evolution as dust opacities become
lower at UV wavelengths.
In heavily irradiated disk atmospheres many species will exist in excited
(ro-)vibrational states, which may then react differently with other species and require addition of state-to-state processes in
the models \citep{Pierce_AHearn10a}.

An accurately calculated UV spectrum including Ly$\alpha$ resonant scattering is
required to calculate photodissociation and photoionization rates, and shielding factors for CO, H$_2$, and
H$_2$O \citep{Bethell_Bergin09,Bethell_Bergin11}.
\citet{Fogel_ea11} adopt their Ly $\alpha$ and UV continuum radiation
field to time-dependent chemistry in disks with grain settling to show
that the Ly $\alpha$ radiation impacts the column densities of HCN,
NH$_3$, and CH$_4$, especially when grain settling is significant.

\citet{Walsh_ea12} studied the impact of photochemistry and X-ray
ionization on the molecular composition in a disk to show that
detailed treatment of X-ray ionization mainly affects N$_2$H$^+$
distribution, while the shape of the UV spectrum affects distribution of
many molecules in the outer disk.

Photodesorption of molecules from icy mantles on grains
is also an important mechanism in the surface and
middle layer of protoplanetary disks. The photodesorption rates were
recently derived experimentally for CO, H$_2$O, CH$_4$, and NH$_3$
\citep{Oeberg_ea07,Oeberg_ea09a,Oeberg_ea09b,Fayolle_ea11a}.
\citet{Walsh_ea10} showed by a chemical model in a protoplanetary disk that
photodesorption significantly affects molecular column densities of
HCN, CN, CS, H$_2$O and CO$_2$ \citep[see also][]{Willacy_07,Semenov_Wiebe11a}.


\bigskip
\noindent
\textit{4.3.2 Chemistry of Organic Molecules}
\bigskip

Although various complex organic molecules, such as glycoaldehyde and
cyanomethanimine \citep[e.g.][]{2013ApJ...765L..10Z}, have been detected
towards luminous
hot cores and hot corinos, H$_2$CO, HC$_3$N \citep{Chapillon_ea12b}, and  c-C$_3$H$_2$ \citep{2013ApJ...765L..14Q}
are the
largest molecules detected towards less luminous protoplanetary disks so
far. On the other
hand, more complex
organic molecules have been found in our Solar System by infrared and
radio observations towards comets \citep[e.g.][]{2011ARA&A..49..471M}, in a
sample return mission from the comet 81P/Wild2 \citep[STARDUST,][]{Elsila_ea09},
and in meteorites. It is an interesting topic to investigate how
these complex molecules were formed in the early Solar System.

Since the detection of mid-infrared lines of small organic molecules and
water by Spitzer Space Telescope \citep[e.g.][]{Carr_Najita08}, they have
been treated in many chemical models of protoplanetary disks
\citep[e.g.][]{2009ApJ...693.1360W,Nomura_ea09,2009A&A...501L...5W}.
\citet{Agundez_ea08} pointed out an importance of gas phase synthesis
with molecular hydrogen to form carbon-bearing organic molecules, while
\citet{Glassgold_ea09} indicated that water abundance is significantly
affected by molecular hydrogen formation rate from which water is mainly
formed through gas-phase reaction in warm disk atmosphere.
\citet{Heinzeller_ea11} showed that transport of molecular hydrogen from
UV-photon-shielded molecular-rich layer to hot disk surface by
turbulent mixing enhances abundances of water and small organic molecules.

Meanwhile, more complex organic molecules, which have not been detected
towards protoplanetary disks yet, are believed to form on grain surfaces
\citep[e.g.][]{Herbst_vanDishoeck09}. \citet{Walsh_ea10} showed that grain
surface chemistry together with photodesorption enhances gas-phase
complex organic molecules, such as HCOOH, CH$_3$OH, and HCOOCH$_3$ in the middle layer
of the outer disk. Also, \citet{Semenov_Wiebe11a} showed that grain
surface chemistry and then organic molecule formation are affected by
turbulent mixing. \citet{Walsh_ea13b} treated further complex grain
surface chemistry (following \citet{Garrod_ea06} and \citet{2008ApJ...682..283G})
which includes heavy-radical recombination reactions efficient on warm grains together
with photodesorption, reactive desorption, and photodissociation of ice
on grains. They showed that complex organic molecules are formed by
the surface reactions on warm grains near the midplane at the disk
radius of $\geq$20AU. The resulting grain-surface fractional abundances
are consistent with those observed towards comets. Also, they predict
based on model calculations of transition lines that strong methanol
lines will be excellent candidates for ALMA observations towards disks.

\bigskip
\noindent
\textbf{4.4 Sensitivity Analysis and intrinsic uncertainties of chemical Models}
\bigskip

To improve our knowledge of the physics and chemistry of protoplanetary
disks, it is necessary to understand the robustness of the chemical models
to the uncertainties in the reaction rate coefficients, which define the
individual reaction
efficiency. Sensitivity methods have been developed for such purposes
and applied to several types of sources
\citep{2004AstL...30..566V,2005A&A...444..883W,2010A&A...517A..21W}.
Those methods follow the propagation of the
uncertainties during the calculation of the abundance species. Error
bars on the computed abundances can be determined and "key" reactions
can be identified. \cite{Vasyunin_ea08} applied such methods for
protoplanetary disk chemistry,
showing that the dispersion in the
molecular column densities is within an order of magnitude at $10^6$ yrs
when chemical reactions are in a quasi-equilibrium state. Abundances of
some molecules could be very sensitive to rate uncertainties, especially
those of X-ray ionization rates in specific regions.

Results of model calculations of chemical reactions also depend on the
applied chemical network in some degree.
Chemical networks used for disk chemistry are usually derived from the
public networks UMIST \citep[][http://www.udfa.net/]{2013A&A...550A..36M})
and OSU
(http://www.physics.ohio-state.edu/$\sim$eric/research.html). Those two
networks have been developed and updated over the years since the
1990s. The latest version of the OSU network has recently been published
under the name of kida.uva.2011 \citep{KIDA}, being
updated from the online database KIDA
(http://kida.obs.u-bordeaux1.fr/). The UMIST network contains a number
of chemical reactions with activation barriers, and even 3-body assisted
reactions, that can become important in the inner regions of the
disks. The OSU/kida.uva network is primarily designed for the outer
parts of the disk were the temperatures stay below 300~K. Additions
have been done to the OSU network by
\cite{Harada_ea10,2012ApJ...756..104H} to use it up to
temperatures of about 800~K.
In \citet{KIDA}, differences in the resulting molecular abundances
computed by different chemical networks were studied in a dense cloud
model to show that errors are typically within an order of magnitude.
It is worth noticing that these networks
contain approximate values for the  gas-phase molecular photodissociation rates
which are valid for the UV interstellar radiation field only. Photorates for different star
spectra have been published by \citet{vDea_06} \citep[see also
][]{Visser_ea09b}.

Contrary to the gas-phase, there is up to
now no detailed database of surface reactions for the interstellar
medium. Only a few networks have been put online by authors. Some of
them can be found on the KIDA website
(http://kida.obs.u-bordeaux1.fr/models).

\bigskip
\centerline{\textbf{5. TOWARDS THE ALMA ERA}}
\bigskip

\bigskip
\noindent
\textbf{5.1 Measuring the turbulence}
\bigskip

Turbulence is a key ingredient for the evolution and composition of disks.

On one hand, the turbulence strength, more precisely the $\alpha$ viscosity parameter,
can be derived indirectly through its impact on disk spreading \citep{Guilloteau_ea11a} and dust settling
\citep{2012A&A...539A...9M}. From the (dust) disk shapes and sizes at mm wavelengths (which sample the disk midplane),
\citet{Guilloteau_ea11a} found
$\alpha \simeq 10^{-3} - 10^{-2}$ at a characteristic radius of 100 AU, {\dbf possibly decreasing} with time.
IR SED analyses, which sample the top layers and are more sensitive to 10~AU scales, provided similar values, although the
derived $\alpha$ depends on the assumed grain sizes and gas-to-dust ratio \citep{2012A&A...539A...9M}.

On another hand, high resolution spectroscopy can directly probe the magnitude of turbulence, by
separating the bulk motions (Keplerian rotation) from local broadening
and isolating the thermal and turbulent contributions to this broadening.
For spatially unresolved spectra, turbulent broadening can smear out the
double-peaked profiles expected for Keplerian disk. This property was used
by \citet{2004ApJ...603..213C}, 
who found supersonic turbulence from CO overtone
bands in SVS13, but lower values from H$_2$O lines, suggesting a highly
turbulent disk surface in the inner few AUs.

With spatially resolved spectro-imaging, the Keplerian motions lead to
iso-velocity regions which follow an arc shape, given by
\begin{eqnarray}
r(\theta) &=& ( GM_{*}/V_\mathrm{obs}^2 ) \sin^2{i}\cos^2{\theta} .
\end{eqnarray}
The dynamical mass and disk inclination control the curvature, while local
broadening controls the width of the emission across the arc. Thus, local
broadening is easily separated from the Keplerian shear, and can be
recovered usually through a global fit of a (semi-)parametric disk model.
Using the PdBI, local line widths 
of order $0.05-0.15$~km/s were found in the early work of
\citet{GD98} 
using CO observations of DM Tau, and
subsequently for LkCa15 and MWC\,480 by \citet{Pietu_ea07}.  Work by
\citet{2004ApJ...616L..11Q} 
found that a smaller linewidth in the range 0.05-0.1\,km/s
was required to reproduce CO J=3-2 observations of the TW Hya disk.  Using CO
isotopologues, \citet{DDG03} 
suggested the turbulence
could be smaller in the upper layers of the disk, a result not confirmed by
\citet{Pietu_ea07} for LkCa15 and MWC\,480 \citep[see also][for GM Aur]{2008A&A...490L..15D}.
Similar values were reported from other molecular
tracers such as CN, HCN \citep{Chapillon_ea12a}, or HCO$^+$
\citep{Pietu_ea07}. The inferred turbulent velocities are subsonic, Mach $\simeq 0.1-0.5$,
which corresponds to viscosity $\alpha$ values of $\simeq 0.01-0.1$ \citep{Cuzzi_ea01}.

These initial measurements suffered from several limitations: one is
sensitivity, resulting in only CO being observed at high S/N, another is the
spectral resolution, and the third is the limited knowledge of the thermal
line width.  Several authors, including \citet{Pietu_ea07},
\citet{2007A&A...469..213I}, 
and \citet{Qi_ea08} have particularly noted
the difficulty of meaningfully constraining turbulent linewidths whose
magnitudes are comparable to the spectral resolution of the data.
\citet{Hughes_ea11a} utilized the higher spectral resolution provided by
the SMA correlator to measure the non-thermal broadening of the CO lines in
TW Hya and HD\,163296.  They used a two independent thermal structures to
derive consistent estimates of turbulent broadening that achieve consistent
results: one derived from modeling the SED of the stars to estimate the
temperature in the CO emitting layer, and one freely variable parametric
model.  Their results indicate a very small level of turbulence in TW\,Hya,
$ < 0.04\,\kms$, but a substantially higher value in HD\,163296 ($ 0.3\,\kms$).


Because of its high optical depth, CO unfortunately only samples a
very thin layer high up in the disk atmosphere.  Measurements with other,
optically thinner, tracers are essential to probe the turbulence across
the disk. However, the distribution of molecules like CN is insufficiently
well understood to remove the thermal component.  For CN (and also C$_2$H),
the apparent excitation temperatures are found to be low, 10 -- 15 K
\citep{Henning_ea10,Chapillon_ea12a}, 
while the temperature in the
molecular rich layer is expected to be around 20-30 K, high enough to
contribute significantly to the local line width. To minimize the contribution of thermal motions
to the local line width such that the turbulence component can be measured,
\citet{Guilloteau_ea12a} used CS, a relatively
``heavy'' molecule ($\mu = 44$, compared to $25 - 28$ for CN, CCH or CO).
Observing the J=3-2 transition with the IRAM PdBI, and with accurate modeling
of the correlator response, they derive a turbulent width of $0.12 \kms$, and
show this value to be robust against the unknowns related to the CS spatial
distribution (kinetic temperature, location above disk plane). It corresponds
to a Mach number of order 0.5, a value which is only reached at several scale
heights above the disk plane in MHD simulations
\citep[e.g.][]{2006A&A...457..343F,2011ApJ...735..122F},  
while chemical models
predict CS to be around 1-2 scale heights.

All measurements so far {\dbf provide disk-averaged values}. The radial variations
are as yet too poorly constrained: assuming a power law radial distribution
for the turbulent width, \citet{Guilloteau_ea12a} inferred an exponent
$0.38 \pm 0.45$, which can equally accommodate a constant turbulent velocity,
or a constant Mach number.  It is also worth emphasizing that the ``turbulent''
width is an adjustable parameter to catch (to first order) all deviations
around the mean Keplerian motion, as could occur in case of e.g. spiral waves
or disk warps (see for example the complex case of AB
Aur in the papers by \citet{Pietu_ea05} and \citet{Tang_ea12}).

The extremely limited number of sources studied so far precludes any general
conclusion to be drawn from the relation between non-thermal linewidth and,
e.g. disk size (which would be expected due to viscous evolution), stellar
mass (which determines the spectrum of ionizing radiation), or evolutionary
state (transitional or non-transitional).  Substantial progress is expected
with ALMA, which will provide much higher sensitivity, higher spectral
resolution, and higher angular resolution, but also the possibility to use
transitions with different optical depths to probe different  altitudes above
the disk plane. The CS molecule is well suited for this, along with a number
of other rotational lines falling in the ALMA bands.

\bigskip
\noindent
\textbf{5.2 Estimating the Ionization and Dead Zone}
\bigskip

One of the most important properties of protoplanetary disks is their ionization structure
which is until now poorly constrained by the observations.
Weakly ionized plasma in rotating configuration is subject to the magnetorotational
instability (MRI) \citep[e.g.][]{MRI}, which is thought to drive turbulence in disks.
Turbulence causes anomalous viscosity of the gas and  thus enables angular momentum transport and
regulates the global disk evolution. It is operational even at very low ionization degree values
of $\la 10^{-10}$ in the inner disk midplane regions completely shielded from the external UV/X-ray radiation
(but rarely from the cosmic ray particles), see also \citet{2013ApJ...777...28C}.
In massive enough disks the inner  region can be shielded even
from the CRPs, either due to magneto-spherical activity of the central star or by a high obscuring
gas column of $\ga 100$~g\,cm$^{-3}$ \citep{un1980,1996ApJ...457..355G,2013ApJ...772....5C}. This leads to the formation of
a ``dead zone'' with damped turbulence, through which transport of matter is severely reduced
\citep[see, e.g.][]{sano,Red2,2012ApJ...761...95F,2012MNRAS.420.3139M,2013ApJ...764...65M,Dzyurkevich_ea13a}.
This has implications on the efficiency of grain growth and their radial migration
\citep{2010ApJ...710L.167H,2012A&A...545A.134M},
planet formation \citep{2007ApJ...670..805O},
disk chemistry and physics \citep{2010ApJ...718.1289S,2011ApJ...740..118L,2012ApJ...744..144F},
development of spiral waves, gaps, and other asymmetric structures
\citep{2010A&A...516A..31M,2012MNRAS.420.2851M,2012MNRAS.422.2399M,2012MNRAS.422.1140G}.

Even with modern computational facilities, the complex interplay between weakly charged plasma
and magnetic fields in disks is very difficult to simulate in full glory of detail.
\citet{Ilgner_Nelson06a,Ilgner_Nelson06b} investigated in detail ionization
chemistry in disks and its sensitivity to various
physical and chemical effects at $r<10$~AU: (1) the X-ray flares from
the young T~Tauri star \citep{Ilgner_Nelson06b}, and (2) the vertical turbulent mixing transport
and the amount of gas-phase metals \citep{Ilgner_Nelson06a}. Using an $\alpha$-disk model with stellar
X-ray-irradiation, they demonstrated that the simple network from \citet{OD74}, using 5 species only, tends to overestimate
the ionization degree since metals exchange charges with polyatomic species, and that magnetically
decoupled ``dead'' region may exist in disks as long as small grains and metals are removed from gas.
The vertical diffusion has no effect on the electron abundances when the metals are absent in the gas because
recombination timescales for polyatomic ions are rapid.  When the dominant ions are metals, the characteristic ionization
timescale becomes so long that the ionization chemistry becomes sensitive to transport,  which drastically reduces the
size of the ``dead'' zone. In the disk model with sporadic X-ray flares (by up to a factor of 100 in the luminosity)
the outer part of the ``dead'' zone disappears, whereas the inner part of the ''dead'' zone evolves along with variations of the
X-ray flux.

The first attempts to model self-consistently disk chemical, physical,
and turbulent structures in full 3D have been
performed by \citet{Turner_ea06} and \citet{Ilgner_ea08}. Both studies
have employed a shearing-box approximation to
calculate a patch of a 3D MHD disk at radii of $\sim 1-5$~AU, treated
the development of the MRI-driven turbulence, and
focused on the multi-fluid evolution of the disk ionization state.

\citet{Ilgner_ea08} have confirmed their earlier findings
\citep{Ilgner_Nelson06,Ilgner_Nelson06a} that turbulent mixing has no
effect on
the disk ionization structure in the absence of the gas-phase metals.
The presence of metals, however, prolongs the
recombination timescale, and the mixing is thus able to enliven the
``dead'' zone at $r\ge5$~AU (with the resulting
$\alpha=1-5\times 10^{-3}$).

The first  3D-MHD disk model coupled with a limited time-dependent ionization chemistry
was made by \citet{Turner_ea07}. They have again confirmed that large-scale turbulent transport
brings charged particles and  radial magnetic fields toward the midplane, resulting in  accretion
stresses there that are
only a few times lower than in the active surface layers.
Later, \citet{2012MNRAS.420.2419F} have performed local 3D radiative MHD simulations of
different radii of a protosolar-like disk, including a simplified chemical network with recombination of charged particles on
dust grains in the presence of the stellar X-rays, cosmic rays and the decay of radionuclides. They have found that
at the distance between 2 to 4~AU a ``dead zone'' appears.

A detailed study of the global gas-grain ionization chemistry in the presence of turbulent mixing for a DM~Tau-like
disk model has been performed by
\citet{Semenov_Wiebe11a}. It was found that the turbulent diffusion does not affect abundances of simple metal ions and a
molecular ions such as C$^+$, Mg$^+$, Fe$^+$, He$^+$, H$_3^+$, CH$_3^+$, NH$_4^+$. On the other hand,
charged species sensitive to the mixing include hydrocarbon ions, electrons, H$^+$, N$_2$H$^+$, HCO$^+$, N$^+$, OH$^+$,
H$_2$O$^+$, etc. They have re-confirmed that the global ionization structure has a layered structure even in the
presence of transport processes: (1) heavily irradiated and ionized, hot
atmosphere, where the dominant ions are C$^+$ and H$^+$, (2) partly UV-shielded, warm molecular layer
where carbon is locked in CO and major ions are H$^+$, HCO$^+$ and H$_3^+$,
and (3) dark, dense and cold midplane, where most of molecules are frozen out onto dust grains, and
the most abundant charged particles are dust grains and H$_3^+$.

\citet{Walsh_ea12} have investigated the impact of photochemistry and X-ray ionization on the molecular composition and
ionization fraction in a T~Tauri disk. The photoreaction rates were calculated using the local UV wavelength spectrum and, the
wavelength-dependent photo cross sections. The same approach was utilized to model the transport of the stellar X-ray radiation.
They have found that  photochemistry is more important for global disk chemistry than the X-ray radiation. More accurate
photochemistry affects the location of the H/H$_2$ and C$^+$/C dissociation fronts in upper disk layer and increases  abundances
of neutral molecules in the outer disk region. \citet{Walsh_ea12} concluded that there is a potential ``dead zone'' with
suppressed accretion located within the inner $\sim 200$~AU midplane region.

Several observational campaigns have begun to test some of these theoretical predictions. \citet{Dutrey_ea07} have performed
high-sensitivity interferometric observations with the IRAM PdBI array of N$_2$H$^+$ 1--0 in the disks around DM~Tau,
LkCa~15, and MWC~480. These data were used to derive the N$_2$H$^+$ column densities and, together with the HCO$^+$ measurements,
were compared with a steady-state disk model with a vertical temperature gradient coupled to a gas-grain chemistry network.
The derived N$_2$H$^+$/HCO$^+$ ratio is on the order of $0.02-0.03$, similar to the value observed in colder and darker
prestellar cores. The observed values qualitatively agree with the modeled column densities at a disk age of
a few million years, but the radial distribution of the molecules were not reproduced. The estimated ionization degree values
from the HCO$^+$ and N$_2$H$^+$ data are $\sim 2-7\times 10^{-9}$ (wrt the total hydrogen density), which are typical
for the warm molecular layers in the outer disk regions.

\citet{2011ApJ...743..152O} have presented interferometric observations with SMA of CO 3--2 in the DM~Tau disk,
and upper limits on H$_2$D$^+$ $1_{1,0}-0_{1,1}$ and N$_2$H$^+$ 4--3. With these data,
IRAM 30-m observations of H$^{13}$CO$^+$ 3--2, and previous SMA
observations of N$_2$H$^+$ 3--2, HCO$^+$ 3--2, and DCO$^+$ 3--2 \citep[see][]{Oeberg_ea10a}, they
constrained ionization fraction using a parametric physical disk model.  They have found that: (1) in a warm molecular layer  ($T
>20$~K) HCO$^+$ is the dominant ion and the ionization degree is $\sim 4 \times 10^{-10}$, (2) in a cooler layer around midplane
where CO is depleted ($\la 15-20$~K) N$_2$H$^+$ and DCO$^+$ are the dominant ions, and  the fractional ionization is $\sim 3
\times 10^{-11}$, and (3) in the cold, dense midplane  ($T <15$~K) the isotopologues of H$_3^+$ are the main charge carriers, and
the fractional ionization drops below $\sim 3\times 10^{-10}$. These observations confirm that the disk ionization degree
decreases towards deeper, denser
and better shielded disk regions.
{\dbf Unfortunately, the best tracer of the ionization in the disk midplane, ortho-H$_2$D$^+$, still remains to be firmly detected
in protoplanetary disks (see Sect.5.3 and Chapter of Ceccarelli et al.).}

\bigskip
\noindent
\textbf{5.3 Deuteration}
\bigskip

Deuterated molecules are important tracers of the thermal and chemical
history in protoplanetary disks and the ISM (see also the chapter by Ceccarelli et al. in this book).  The cosmic D/H
ratio in
the local ISM
is D/H$\approx 1.5\,10^{-5}$ \citep{2006ApJ...647.1106L},
with a major reservoir of D locked in HD in dense regions.
The deuterium fractionation processes redistribute
the elemental deuterium from HD to other species at low
temperatures, enhancing abundances and D/H ratios of many polyatomic species
by orders of magnitude, as observed in the cold ISM
\citep[e.g.][]{2003ApJ...585L..55B,2007ARA&A..45..339B,2013arXiv1310.3151H}.

This is due to the difference in zero point vibrational energy between
the H-bearing and D-bearing species,
which implies barriers for backward reactions and thus favors
production of deuterated species.
The key process for gas-phase fractionation at $T\la 10-30$~K is
deuterium enrichment of H$_3^+$:
H$_3^+$ + HD $\leftrightarrows$ H$_2$D$^+$ + H$_2$ +
232~K \citep{Millarea89,GHR_02}.
Upon fractionation of H$_3^+$, proton exchange reactions
transfer the D enhancement to more complex gaseous species.
One of the key reactions of this kind in protoplanetary disks is the
formation of DCO$^+$:
H$_2$D$^+$ + CO $\leftrightarrows$ DCO$^+$ + H$_2$,
which produces an enhanced DCO$^+$/HCO$^+$ ratios at in outer cold
disk midplanes (at $\sim 10-30$~K).
A similar reaction produces  N$_2$D$^+$ in the coldest regions where
gaseous CO is frozen-out:
H$_2$D$^+$ + N$_2$ $\leftrightarrows$ N$_2$D$^+$ + H$_2$.

In the outer midplane regions where CO and other  molecules are
severely depleted, multiply-deuterated
H$_3^+$ isotopologues attain large abundances \citep{2011arXiv1110.2644A}.
Upon dissociative recombination large amounts of atomic  H and D are produced.
Atomic hydrogen and deuterium can then stick to grain surfaces and
hydrogenate or deuterate
the first generation of relatively simple ices. After that,
CRP-driven/UV-driven photo- and
thermal processing of ice mantles allow more complex (organic)
deuterated ices to be synthesized
(e.g. multiply-deuterated H$_2$O, H$_2$CO, CH$_3$OH, HCOOH, etc.).

Another pathway to redistribute the elemental D from HD via gas-phase
fractionation is
effective at higher temperatures, $\la 70-80$~K:
CH$_3^+$ + HD $\leftrightarrows$ CH$_2$D$^+$ + H$_2$ + 390~K
\citep{Asvany_ea04} and C$_2$H$_2^+$ + HD $\leftrightarrows$
C$_2$HD$^+$ + H$_2$ + 550~K
\citep{Herbst_ea87}. After that, DCN molecules can be produced : N +
CH$_2$D$^+$ $\rightarrow$
DCN$^+$ + H$_2$, followed by protonation of DCN$^+$ by H$_2$ or HD and
 dissociative recombination of
DCNH$^+$ \citep{Roueff_ea07}.

Only three deuterated species have been detected in disks: DCO$^+$, DCN, and HD.
The detection of the ground transition ${1_{0,1}-0_{0,0}}$ of HDO in
absorption in the disk of DM~Tau
by \citet{2005ApJ...631L..81C} was later refuted by
\citet{2006A&A...448L...5G}, 
and remains unproven.
DCO$^+$ was first detected in the TW Hya disk \citep{2003A&A...400L...1V,Qi_ea08}, 
then in DM Tau by \citet{2006A&A...448L...5G}, 
and LkCa15 by \citet{Oeberg_ea10a}.
The interpretation of the apparent [DCO$^+$]/[HCO$^+$] ratio, $\sim 0.02$, is
not simple, as HCO$^+$ is in general optically
thick \citep[e.g.][]{2011ApJ...734...98O},  
and both molecules have different  spatial
distributions \citep{2012ApJ...749..162O}. 

DCN has been detected so far in two disks: in TW~Hya by
\cite{Qi_ea08} and in LkCa~15 by \cite{Oeberg_ea10a}.
The SMA and  ALMA Science Verification observations show different
spatial distributions of DCN and DCO$^+$,
which imply different formation pathways \citep[see][]{Qi_ea08,2012ApJ...749..162O}. 
Deuteration for DCN proceeds more efficiently through proton exchange
with deuterated light hydrocarbons
like CH$_3^+$, whereas for DCO$^+$ low-temperature fractionation via
H$_3^+$ isotopologues is
important (see above).

Recently, the far-infrared fundamental rotational line of HD has been
detected in the TW~Hya disk with {\it
Herschel} \citep{2013Natur.493..644B}. 
The abundance distribution of HD
closely follows that of H$_2$.
Therefore, HD, which has a weak permanent dipole moment, can be used
for probing disk gas mass.
The inferred mass of the TW~Hya disk is $\ga 0.05M_\odot$, which is
surprising for a
$\sim 3-10$Myr-old system.

Among deuterated molecular species, H$_2$D$^+$ is potentially a
critical probe of diskc
midplanes, where it can be a dominant charged
species  \citep{Ceccarelli_Dominik05}.
It is rapidly destroyed by the ion-molecule reactions with CO and
other molecules,
and abundant enough only in the high density cold midplane regions,
where most of the
molecules are frozen onto grains. Its lowest transition is
unfortunately at 372~GHz,
where the atmospheric transparency is limited. Current searches failed
to detect it,
the initial tentative reports of
\citet{2004ApJ...607L..51C} 
 being superseded by 3 times more sensitive
upper limits
of \citet{Chapillon_ea11} using the JCMT and APEX in DM Tau and TW Hya.
\citet{Chapillon_ea11} compared the upper limit with the prediction
from several disk models, varying the density, temperature and outer radius), grain sizes
(0.1, 1 and 10 $\mu$m), CO abundances (10$^{-4}$,10$^{-5}$ and 10$^{-6}$) and rate of cosmic rays ionization (10$^{-17}$,
3.10$^{-17}$
and 10$^{-16}$ s$^{-1}$). The data only firmly exclude cases with both a low CO abundance and small grains.
This study also indicates that H$_2$D$^+$ is more difficult to detect than expected, and that even
with the full ALMA, significant integration times will be needed to probe the physics and chemistry
of the cold depleted midplane of T~Tauri disks through H$_2$D$^+$ imaging.

\bigskip
\noindent
\textbf{5.4 Gas in Inner Cavities}
\bigskip

It is worth mentioning that (sub)mm observations are in some cases sensitive
to the gaseous content of the inner disks. The simple way that comes to mind
is the use of observations spatially resolving the inner disk (whatever
arbitrary size an "inner disk" refers to). Unless the inner region under
consideration is large, this requires a very fine angular resolution, which
limits its applicability to the bright disks (and bright lines) and large holes.
However, it is not necessary to spatially resolve the inner parts of the
disk to have access to its content. Since disks are geometrically thin with
a well ordered velocity field, there is a mapping between velocity axis and
position within the disks. Specifically, in a Keplerian disk, the line wings
originate from the inner parts, so e.g. the absence of high-velocity wings
indicates an inner truncation. However the line wings are also fainter than
the core which makes such analysis challenging from a sensitivity point of
view. Another complication arises from the fact this flux density from this
wings is comparable if not weaker to the continuum flux density, calling for
both a very good bandpass calibration and continuum handling.

The archetype of a large inner hole lies at the center of GG\, Tau, the famous
multiple system, with a circumbinary ring surrounding GG\, Tau\, A.
\citet{DGS94,Guilloteau_ea99} found that the dust and the gas circumbinary ring
plus disk had the same inner radius.  \citet{Guilloteau_ea01} later found
$^{12}$CO 2--1 emission within the ring that possibly traces streaming
material from the circumbinary ring feeding the circumprimary disks that
would otherwise disappear through accretion of their material onto the
host stars.

A handful of other systems have been imaged at high angular resolution
in molecular lines, providing interesting complementary information to that
of the continuum (see Chapter by Espaillat et al.). For example,
\citet{Pietu_ea07} found from CO 2--1 observations that CO is present in the
inner part of LkCa\,15 disk where dust emission present a drop in emissivity a
situation similar to the AB\,Aur system \citep{Pietu_ea06,Tang_ea12}.

Using a spectroscopic approach, \citet{2008A&A...490L..15D} 
analyzed the J=2-1 line
emission of CO isotopologues in the GM Aur disk (PdBI data), and used
a global fit to the data to
overcome the sensitivity issue. They could show that the inner radius in
CO was comparable to that of the dust \citep{Hughes_ea09}.  A similar
method was used to infer the presence of an
inner gas cavity in the 40\,Myr-old 49 Ceti system \citep{2008ApJ...681..626H}. 
More recently, \citet{2012ApJ...757..129R} 
performed a similar analysis on
TW\,Hya taking advantage of the brightness of the relatively close system
and enhanced sensitivity of the ALMA array. This allowed them to not only
trace CO down to about 2\,AU from the central star, but also to point out
that simple model fails to reproduce the observed line wings calling for
alternative models such as, e.g. a warped inner disk.

ALMA is already providing examples of the kind of structures that may be
present in the inner disk, by the analogous examples at larger
radii. Using ALMA, \citet{2013Natur.493..191C} were able to image multiple molecular
gas species within the dust inner hole of HD\,142527 and utilize the varying
optical depths of the different species to detect a flow of gas crossing the
gap opened by a planet while \cite{2013Sci...340.1199V} recently imaged a dust
trap in the disk surrounding Oph IRS 48. Over the next several years,
ALMA will likely uncover similar structures, and many more
unexpected ones, at radii between a few and 30 AU.

\bigskip
\centerline{\textbf{5. SUMMARY AND OUTLOOK}}
\bigskip

Since PPV, thermo-chemical models of disks, as chemical models, have been significantly
improved but the detection rate of new molecules in disks has not significantly increased.
However, there are some trends which  appear from the recent observational results.
Among them, it is important to mention:

\begin{itemize}

\item So far, only simple molecules have been detected, the most complex being HC$_3$N and c-C$_3$H$_2$, showing
that progress in understanding the molecular complexity in disks will require substantial observational efforts,
even with ALMA. Similarly, the upper limits obtained on H$_2$D$^+$ indicate that probing the disk midplane
will be difficult.

\item Multi-line, multi-isotopologue CO interferometric studies allow observers and modelers to
retrieve the basic density and temperature structures of disks. This method will become more and
more accurate with ALMA.

\item {\bbf The} CO, CN, HCN, CS and CCH molecular studies apparently reveal low temperatures in the
gas phase in several T~Tauri disks. These observations suggest that the molecular layer is at least partly
colder than predicted and that there are some ingredients still missing in our understanding of
the disk physics and/or chemistry. More accurate ALMA data will refine these studies.

\item Dust is one of the major agent shaping the disk physics and chemistry. In particular, it plays
a fundamental role to control the UV disk illumination and hence the temperature. It also strongly
influences the chemistry since a significant part of the disk is at a temperature which is below
the freeze out temperature of most molecules. Although substantial progress has been
made in the chemical modeling, a more comprehensive treatment of chemical reactions on grain surface remains
to be incorporated.

\item Deriving the total amount of gas mass is the most difficult task
because the bulk of the gas, H$_2$, cannot be directly traced.
The recent detection of HD with Herschel is therefore very promising.
In the near future, GREAT on SOFIA may allow detection of HD lines in other disks.
Probing the gas-to-dust ratio will be then an even more complex step,
as recent dust observations reveal that the dust properties also changes
radially. ALMA will fortunately provide a better accuracy on the observable
aspect of dust properties, the emissivity index.

\item With an adequate angular resolution and sensitivity, ALMA will provide the first
images of inner dust and gas disks ($R\leq 30$~AU), revealing the physics and chemistry
of the inner part. In particular, these new observations will provide the first
quantitative constrain on the ionization status, the turbulence and the kinematics in the planet
forming area, as any departure to 2D-symetry.

\end{itemize}

\vspace{1.0cm}
\textbf{Acknowledgments.}
Anne Dutrey would like to acknowledge all CID members for a very fruitful collaboration, in particular Th.Henning, R.Launhardt,
F.Gueth and K.Schreyer. We also thank all the KIDA team (http://kida.obs.u-bordeaux1.fr) for providing chemical
reaction rates for astrophysics. Anne Dutrey and St\'ephane Guilloteau thank the French Program PCMI for providing
financial support. Dmitry Semenov acknowledges support by the Deutsche Forschungsgemeinschaft through SPP 1385:
The first ten million years of the solar system - a planetary materials approach (SE 1962/1-2 and 1962/1-3).

\bigskip

\bibliography{final}
\bibliographystyle{ppvi_lim1}

\end{document}